\documentclass[showpacs,twocolumn,prb]{revtex4-1}
\usepackage{graphicx}
\graphicspath{{figures/}} % Location of the graphics files
\newcommand{\bwt}{\begin{widetext}}
\newcommand{\ewt}{\end{widetext}}

\newcommand{\be}{\begin{equation}}
\newcommand{\ee}{\end{equation}}
\def\bea {\begin{eqnarray}}
\def\eea {\end{eqnarray}}  
\def \bk{{\bf k}}
\def \bkp{{\bf k^\prime}}

\def \iwm {i\omega_m}
\def \iwmp {i\omega_{m^\prime}}
\def \kiwm {\bk,i\omega_m}

\usepackage{hyperref}
\hypersetup{
    colorlinks=true,
    linkcolor=blue,
    filecolor=magenta,
    urlcolor=blue,
    citecolor=magenta
}

\usepackage{color} 
\newcommand{\itt}{\it}
\newcommand{\black}{\textcolor{black}}

\usepackage{comment}
\def\comment#1{}
\newcommand{\com}{\black}

\usepackage{ulem}   % to strike things out
\begin{document}

\title{Eliashberg Theory: a short review}

\author{F. Marsiglio}
\affiliation{Department of Physics, University of Alberta, Edmonton, AB, Canada T6G~2E1}

\begin{abstract}
Eliashberg theory is a theory of superconductivity that describes the role of phonons in providing the attractive interaction between two electrons.
Phonon dynamics are
taken into account, thus giving rise to retardation effects that impact the electrons, in the form of a frequency-dependent electron self-energy.
In the superconducting state, this means that the order parameter, generally considered to be a static quantity in the Bardeen-Cooper-Schrieffer (BCS) theory, also
becomes frequency dependent. Here we review the finite temperature formulation of Eliashberg theory, both on the imaginary and real
frequency axis, and briefly display some examples of the consequences of a dynamical, as opposed to static, interaction. Along the way
we point out where further work is required, concerning the validity of some of the assumptions used.
\end{abstract}

\pacs{}
\date{\today }
\maketitle

\section{introduction}

Superconductivity is a remarkable phenomenon, not least because it represents a manifestation of the quantum world on a macroscopic scale.
It is spectacularly demonstrated with levitating train sets,\cite{levitating_trainset_cornell} and indeed this property and many others of superconductors
are slowly being utilized in everyday applications.\cite{maglev} However, the established practice of incorporating superconductors into the real world
should not be taken as an indication that ``the last nail in the coffin [of superconductivity]''\cite{parks_preface} has been achieved. On the contrary, 
in the intervening half-century since this quote was written, many new superconductors have been discovered, and we have reached a point 
where it is clear that a deep lack of understanding\cite{lack} of superconductivity currently exists. 
Reference [\onlinecite{physicac2015}] compiles a series of articles reviewing the various ``families'' or classes of superconductors, where
one can readily see common and different characteristics. At the moment many of these classes require a class-specific mechanism for superconductivity,
a clearly untenable situation, in my opinion. Further classes have been discovered or expanded upon since, such as nickelates,\cite{li19}  and the hydrides 
under pressure,\cite{kong19} for example.

Our ``deep lack of understanding'' should not be taken to indicate that theoretical contributions have not been forthcoming. In fact there have been remarkable
contributions to key theoretical ideas in physics that stem from research in superconductivity, starting with London theory\cite{london50} and 
Ginzburg-Landau\cite{ginzburg50} theory, through to BCS theory.\cite{bardeen57} When Gor'kov\cite{gorkov58} recast the BCS theory of superconductivity
in the language of Green functions, then the stage was set for Eliashberg\cite{eliashberg60a,eliashberg60b} to
formulate the theory that bears his name. It is fitting that we honour the lasting impact of his work with this brief review, on the occasion of his 90th birthday,
and the 60th anniversary of the publication of two papers that paved the path for considerable future quantitative work in superconductivity. Based on an index I
am fond of using for famous people, his name appears in titles of papers 248 times, and in abstracts and keywords 1439 times.\cite{web_elias} 

Before proceeding further, we wish to make some remarks about the nature of this review. It will necessarily repeat material from previous reviews, which
we catalogue as follows. Scalapino\cite{scalapino69} and McMillan and Rowell\cite{mcmillan69} perhaps gave one of the first comprehensive reviews of
both calculations and experiments that provide remarkable evidence for the validity of Eliashberg theory for various superconductors. 
These reviews were provided in the comprehensive monogram
by Parks;\cite{parks69} the reader should refer to this monogram and the references therein, as we cannot possibly properly 
reference all the primary literature sources before ``Parks'', as this would consume too many pages here.
The author list in Parks is a who's who of experts in superconductivity, with two notable exceptions, John Bardeen and Gerasim (Sima) Eliashberg. 

A subsequent very influential review was that of Allen and Mitrovi\'c,\cite{allen82} where mostly superconducting $T_c$ was discussed.
These authors highlighted the expediency of doing many calculations on the imaginary frequency axis, a possibility first noted in Ref.~[\onlinecite{owen71}] and
utilized to great advantage in subsequent years.\cite{bergmann73,rainer74,allen75,daams81} 

A few years later Rainer wrote a ``state-of-the-union'' address\cite{rainer86} on first principles calculations of superconducting $T_c$ in which a challenge was
issued to both band structure and many-body theorists. For the former, the missing ingredient was a {\it complete} (italics are mine) calculation of the
electron-phonon coupling. These were first calculated in the 1960's (e.g. Ref.~[\onlinecite{carbotte68}]) but have experienced vast improvement over the
past 50 years, through the adoption and improvement of Density Functional Theory methods, plus the increased computational ability achieved in the
intervening decades. Excellent summaries of this progress is provided in Refs.~[\onlinecite{boeri18},\onlinecite{sanna17},\onlinecite{giustino17}], where two alternative
procedures are described. The first follows the original route of determining the electron-phonon interaction and including this as input to the Eliashberg equations,
while the second aims to treat both the electron-phonon-induced electron-electron interaction {\it and} the direct Coulomb interaction on an equal footing. The
result is advertised to meet Rainer's challenge and calculate $T_c$ and other superconducting properties without any experimental input. It is noteworthy that
in the second formulation (see also Refs.~[\onlinecite{luders05}] and [\onlinecite{marques05}]) the equations are BCS-like and do {\it not} depend on frequency, but
only momentum. In the usual formulation, the calculation of $\mu^\ast$, the effective direct electron-electron Coulomb interaction, is often simply assigned a (small)
numerical value, and therefore is treated phenomenologically as a fitting parameter. The challenge to many-body theorists remains, as more superconductors have
been discovered that seem to extend beyond the weak coupling regime, and likely require descriptions beyond BCS and Eliashberg theory.

In 1990 the review by Carbotte\cite{carbotte90} provided a comprehensive update for a number of thermodynamic properties of various superconductors
known at the time, including the high temperature cuprate materials. Partly for this reason his review is titled ``{\it Properties of boson-exchange superconductors},''
since there was a feeling at the time (and still is in parts of the community) that the Eliashberg framework might apply to these superconductors, but with exchange of a boson other
than the phonon. Quite a few years later we wrote a review jointly,\cite{marsiglio08} focusing on ``{\it electron-phonon superconductivity}.'' This review summarized known 
properties and extended results to dynamical properties such as the optical conductivity, building on earlier work in Ref.~[\onlinecite{lee89}] and mini-reviews
in Ref.~[\onlinecite{marsiglio97}]. More recently Ummarino has published a mini-review with some generalizations to multi-band and the iron pnictide 
superconductors.\cite{ummarino13}

The other remark we should make is that while Eliashberg theory has been extremely successful, we will also point out the limitations that exist. Indeed, these were recognized
right from the beginning, with both Eliashberg\cite{eliashberg60a} and Migdal\cite{migdal58} emphasizing that limitations exist on the value of the dimensionless coupling parameter, $\lambda$, due to the expected phonon softening that would occur as $\lambda$ increases. They claimed an upper limit of $\lambda \approx 1$, which then
significantly restricts the domain of validity of the theory. Constraints on the parameters would be a constant theme over the ensuing years. In 1968 McMillan\cite{mcmillan68}
gave more quantitative arguments for a maximum $T_c$, based on the expected relationship between the coupling strength and the phonon frequency. 
This was reinforced by Cohen and Anderson\cite{cohen72} and has been discussed critically a number of times since.\cite{moussa06,esterlis18}
Alexandrov\cite{alexandrov01} has also raised objections, based on polaron collapse, a topic we will revisit later.

Some of the early history regarding the origins of the electron-phonon interaction was provided in Ref.~[\onlinecite{marsiglio08}] and will be omitted here. By the early
to mid 1950's Fr\"ohlich\cite{frohlich54} and Bardeen and Pines\cite{bardeen55} had established that the effective Hamiltonian for the electron-phonon interaction had a
potential interaction of the form\cite{ashcroft76}
\begin{equation}
V_{\bf k,k^\prime}^{\rm eff} = {4 \pi e^2 \over ({\bf k-k^\prime})^2 +
k_{\rm TF}^2}
\biggl[
1 + {\hbar^2 \omega^2({\bf k-k^\prime}) \over
({\epsilon_{\bf k} - \epsilon_{\bf k^\prime}})^2  -
\hbar^2 \omega^2({\bf k-k^\prime}) }
\biggr],
\label{bardeen_pines}
\end{equation}
where $k_{\rm TF}$ is the Thomas--Fermi wave
vector, and $\omega({\bf q})$
is the dressed phonon frequency. This part of the Hamiltonian represents the pairing interaction between two electrons with wave vectors  ${\bf k}$  and  ${\bf k^\prime}$
in the First Brillouin Zone (FBZ) and energies $\epsilon_{\bf k}$ and $\epsilon_{\bf k^\prime}$. The interaction Hamiltonian written in this form is often said to have
``the phonons integrated out.'' It was on the basis of this Hamiltonian that Bardeen, Cooper and Schrieffer
(BCS)\cite{bardeen57} formulated a model Hamiltonian with an attractive (negative in sign) interaction for electron energies near the Fermi energy, $\epsilon_F$.\\

\section{The Eliashberg Equations}

Eliashberg,\cite{eliashberg60a} following what Migdal\cite{migdal58} had calculated in the normal state, did not ``integrate out the phonons,'' but instead adopted the Hamiltonian
\begin{eqnarray}
H  & =  & \sum_{{\bf k} \sigma}(\epsilon_{\bf k} - \mu) c^\dagger_{{\bf k}
\sigma} c_{{\bf k} \sigma} + \sum_{\bf q} \hbar \omega_{\bf q}
a^\dagger_{\bf q} a_{\bf q} \nonumber \\ & & + {1 \over \sqrt{N}}
\sum_{{\bf k} {\bf k}^\prime \atop \sigma} g({\bf k},{\bf
k}^\prime) \bigl(a_{{\bf k} - {\bf k}^\prime} + a^\dagger_{-({\bf
k} - {\bf k}^\prime)} \bigr) c^\dagger_{{\bf k}^\prime \sigma}
c_{{\bf k} \sigma}\quad . \label{ham_BKF_mom}
\end{eqnarray}
where $c_{{\bf k}\sigma}$ ($c^\dagger_{{\bf k}\sigma}$) is the annihilation (creation) operator for an electron with spin $\sigma$ and wave vector ${\bf k}$,
and $a_{\bf q}$ ($a^\dagger_{\bf q}$) is the annihilation (creation) operator for a phonon with wave vector ${\bf q}$. The electron-phonon coupling function,
$g({\bf k},{\bf k^\prime})$ is generally a function of both wave vectors (and not just their difference), and in principle is calculable with the Density Functional Theory
Methods mentioned earlier. Very often models are adopted based on simple (e.g. tight-binding) considerations.

Eliashberg then applied the apparatus of field theory to formulate a pairing theory that accounts for the dynamics of the interaction, i.e. for retardation effects.
A sketch of the derivation, taken from Rickayzen\cite{rickayzen65} (see also Ref.~[\onlinecite{marsiglio08}]), is provided in the Appendix. This is my favourite
derivation, as it does not rely on a formalism (e.g. the Nambu formalism) whose validity requires an act of faith (or, you simply work through everything anyways, to
ensure that the formalism ``works''). 

Following Eliashberg\cite{eliashberg60b} with more modern notation,\cite{scalapino69,allen82}
the ``normal'' self energy $\Sigma({\bf k},i\omega_m)$ is separated out into even and odd (in Matsubara frequency) parts, so that two new functions,
$Z$ and $\chi$ are defined:
\begin{eqnarray}
i\omega_m \bigl[ 1 - Z({\bf k},i\omega_m) \bigr]  & \equiv &
{1 \over 2} \bigl[ \Sigma(\kiwm) - \Sigma({\bf k},-i\omega_m) \bigr]
\nonumber \\
\chi(\kiwm) & \equiv & 
{1 \over 2} \bigl[ \Sigma(\kiwm) + \Sigma({\bf k},-i\omega_m) \bigr].
\label{even_odd}
\end{eqnarray}

The equations that emerge are
\bwt

\bea
Z({\bf k},i\omega_m) = 1 + {1 \over N\beta}
\sum_{\bkp,m^\prime}
{\lambda_{\bk \bkp}(\iwm - \iwmp) \over g({\epsilon_F})}
{ \bigl(\omega_{m^\prime} / \omega_m \bigr)
Z({\bf k^\prime},i\omega_{m^\prime}) \over 
\omega_{m^\prime}^2 Z^2({\bf k^\prime},i\omega_{m^\prime}) + 
\bigl( \epsilon_{\bf k^\prime} - \mu + \chi({\bf k^\prime},i\omega_{m^\prime})
\bigr)^2  + \phi^2({\bf k^\prime},i\omega_{m^\prime})}
\label{ga1}
\\                                                          
\chi({\bf k},i\omega_m) = - {1 \over N\beta}
\sum_{\bkp,m^\prime}
{\lambda_{\bk \bkp}(\iwm - \iwmp)  \over g({\epsilon_F})} 
{ \epsilon_{\bf k^\prime} - \mu + \chi({\bf k^\prime},i\omega_{m^\prime}) 
\over
\omega_{m^\prime}^2 Z^2({\bf k^\prime},i\omega_{m^\prime}) +
\bigl( \epsilon_{\bf k^\prime} - \mu + \chi({\bf k^\prime},i\omega_{m^\prime})
\bigr)^2  + \phi^2({\bf k^\prime},i\omega_{m^\prime})}
\label{ga2}   
\eea
along with the equation for the order parameter: %gap equation (Eq. (\ref{g2})):
\be
\phi({\bf k},i\omega_m) = {1 \over N\beta}
\sum_{\bkp,m^\prime}
\bigl[ {\lambda_{\bk \bkp}(\iwm - \iwmp)  \over g({\epsilon_F})} - V_{\bf k, k^\prime} \bigr] 
{ \phi({\bf k^\prime},i\omega_{m^\prime})
\over
\omega_{m^\prime}^2 Z^2({\bf k^\prime},i\omega_{m^\prime}) +
\bigl( \epsilon_{\bf k^\prime} - \mu + \chi({\bf k^\prime},i\omega_{m^\prime})
\bigr)^2 + \phi^2({\bf k^\prime},i\omega_{m^\prime})}.
\label{ga3}     
\ee
These are supplemented with the electron number equation, which determines the chemical potential, $\mu$:
\bea
n_e & = & 1 - {2 \over N\beta} \sum_{\bkp,m^\prime} 
{ \epsilon_{\bf k^\prime} - \mu + \chi({\bf k^\prime},i\omega_{m^\prime})
\over
\omega_{m^\prime}^2 Z^2({\bf k^\prime},i\omega_{m^\prime}) +
\bigl( \epsilon_{\bf k^\prime} - \mu + \chi(({\bf k^\prime},i\omega_{m^\prime})
\bigr)^2  + \phi^2({\bf k^\prime},i\omega_{m^\prime})}.
\label{ga4}
\eea

\ewt

Written in this way both $Z$ and $\chi$ are even functions of $i\omega_m$ (and, as
we've assumed from the beginning, they are also even functions of ${\bf k}$). With electron-phonon pairing the anomalous self energy,
which determines the anomalous pairing amplitude $\phi({\bf k}, i\omega_m)$, is
also an even function of Matsubara frequency. A generalization of this latter result, giving rise to so-called Berezinskii\cite{berezinskii74} ``odd-frequency'' pairing, is
beyond the scope of this review. A survey of Berezinskii pairing is given in Ref.~[\onlinecite{linder19}]. 

Other symbols in Eqs.~(\ref{ga1}-\ref{ga4}) are as follows.% Eqs.~(\ref{ga1},\ref{ga2},\ref{ga3},\ref{ga4}) are as follows.
The number of lattice sites is given by $N$, the parameter $\beta \equiv 1/(k_BT)$, where $k_B$ is the Boltzmann constant and $T$ is the temperature,
$\mu$ is the chemical potential, and $g({\epsilon_F})$ is the electronic density of states at the Fermi level
in the band. 
These equations are generally valid for multi-band systems, and then the labels ${\bf k}$ and ${\bf k^\prime}$ are to be understood to include band indices. However,
we shall proceed for simplicity with the assumption of a single band, with single particle energy $\epsilon_{\bf k}$.
Because we are assuming finite temperature right from the start, the equations are written on
the imaginary frequency axis, and are functions of the Fermion Matsubara frequencies, $i\omega_m \equiv \pi k_B T (2m-1)$, with $m$
an integer. Similarly the Boson Matsubara frequencies are given by $i\nu_n \equiv 2 \pi k_B T n$, where $n$ is an integer.
Finally, we have also included a direct Coulomb repulsion in the form of $V_{\bf k, k^\prime}$, which in principle represents the full (albeit screened) Coulomb
interaction between two electrons. 

The key ingredient of Eliashberg theory (as opposed to BCS theory) is the presence of the electron-phonon propagator, contained in
\be
\lambda_{\bk \bkp} (z) \equiv \int_0^\infty {2 \nu \alpha_{\bk \bkp}^2
F(\nu) \over \nu^2 - z^2} d \nu
\label{lambda_z}
\ee
\noindent with $\alpha_{\bk \bkp}^2 F(\nu)$ the spectral function of the phonon Green function. This function is sometimes written as
$\alpha_{\bk \bkp}^2(\nu) F(\nu)$ to emphasize that the coupling part ($\alpha^2$) can have significant frequency dependence. This spectral function is often 
called the Eliashberg function. Equation~(\ref{lambda_z})
has been used as a ``bosonic glue'' to generalize the application of the Eliashberg/BCS formalism to beyond that of phonon exchange. Very often the boson is a collective
mode of the very degrees of freedom that are superconducting, i.e. the conduction electrons. Examples include spin fluctuations or plasmons, but this work
is on more questionable footing.\cite{verga03}

A significant anisotropy may exist, specifically through the nature of the coupling in the Eliashberg function. Since the important physical attribute of the
Eliashberg formalism beyond BCS is retardation, and therefore in the frequency domain, we will nonetheless neglect anisotropy in what follows.\cite{choi02}
More nuanced arguments for the wave vector dependence of the electron-phonon coupling are
provided in Ref.~[\onlinecite{allen82}], connected to the energy scale hierarchy $\epsilon_F >> \nu_{\rm phonon} >> T_c$, where $\nu_{\rm phonon}$ is a typical
phonon energy scale (note that we have adopted the standard practice of dropping $\hbar$ and $k_B$, and therefore we refer to temperatures and phonon frequencies as
energies). Indeed very often the neglect of anisotropy was justified by the study of so-called ``dirty'' superconductors, 
where the presence of impurities served to self-average over
anisotropies. We will also drop the wave vector dependence in the direct Coulomb repulsion, although this step is less justified. It means that the direct Coulomb repulsion
is represented by a single parameter, which we will call $U$, since this is what we would obtain by reducing the long-range Coulomb repulsion with an on-site
Hubbard interaction with strength given by $U$. This is one of the weak points of the Eliashberg description of superconductivity --- an inadequate description of
correlations due to Coulomb interactions. In what follows we will focus more attention on the retardation effects, since this is the part that Eliashberg theory is
best designed to handle properly.

Once we drop the wave vector dependence in the coupling function, all quantities ($Z, \chi$ and $\phi$) become independent of wave vector. The integration over
the First Brillouin Zone can then be performed, although here a series of approximations are utilized. The result can lead to confusion, so we provide some detail
here. First, once it is determined that the unknown functions in Eqs.~(\ref{ga1}-\ref{ga4}) do not depend on wave vector, we can replace the sum over wave vectors in
the first Brillouin zone with an integration over the electronic density of states,
\begin{equation}
{1 \over N} \sum_k \rightarrow \int_{\epsilon_{\rm min}}^{\epsilon_{\rm max}} d \epsilon \ g(\epsilon),
\label{sum_replacement}
\end{equation}
where $g(\epsilon)$ is the single electron density of states and $\epsilon_{\rm min}$ and $\epsilon_{\rm max}$ are the minimum and maximum energies of the
electronic band. Since typically the energy scales are such that $T_c << \nu_{\rm phonon} << \epsilon_F$,
the variation of the electronic density of states away from the Fermi level is of little importance, so the approximation $g(\epsilon) \approx g(\epsilon_F)$ is used.
Exceptions to this case have been discussed previously, and one is referred to Ref.~[\onlinecite{marsiglio08}] for references.
The remaining integration is now elementary and yields a combination of inverse tangent and logarithmic functions.\cite{marsiglio92}

However, the ``standard'' practice is to extend these integrations over energy to $\pm \infty$ (and similarly adopt particle-hole symmetry so $\epsilon_{\rm min} \equiv
-\epsilon_{\rm max}$, along with $\mu = 0$, and now $\epsilon_{\rm max} \rightarrow \infty$) with the philosophy that the remainder of the integrand ensures 
that these additional contributions are negligible. This is in fact {\it not true} and the term proportional to $V_{\bf k, k^\prime}$ in Eq.~(\ref{ga3}) (now replaced 
by $U$ as described above) will result in a Matsubara sum that diverges if this procedure is carried out without thought. 
So in fact the bandwidth parameters are required in the integration for
this term and the resulting inverse tangent function results in a soft cutoff at $\omega_m \approx \epsilon_{\rm max}$. This is most often replaced with a hard (step-function)
cutoff. Because of the assumptions about particle-hole symmetry the function $\chi({\bf k}, i\omega_m)$ in Eq.~(\ref{ga2}) is identically zero, and Eq.~(\ref{ga4}) becomes
meaningless (so the occupation is no longer considered an input parameter).

The more highly simplified equations that result are
\bwt

\be
Z(i\omega_m)  =  1+ {\pi T_c \over \omega_m} \sum_{m^\prime = -\infty}^{+\infty} \lambda(i\omega_m - i\omega_{m^\prime})
{\omega_{m^\prime} Z(i\omega_{m^\prime}) \over \sqrt{\omega^2_{m^\prime} Z^2(i\omega_{m^\prime}) + \phi^2(\omega_{m^\prime})}}
\label{gb1}
\ee
\be
\phi(i\omega_m)  =  \pi T_c \sum_{m^\prime = -\infty}^{+\infty} 
\left[ \lambda(i\omega_m - i\omega_{m^\prime}) - u \ \theta ( {W \over 2} - |\omega_{m^\prime}| ) \right]
{ \phi(i\omega_{m^\prime}) \over \sqrt{\omega^2_{m^\prime} Z^2(i\omega_{m^\prime}) + \phi^2(\omega_{m^\prime})}}.
\label{gb3}
\ee
where $W$ is the bandwidth (i.e. $\epsilon_{\rm max}  = - \epsilon_{\rm min} \equiv W/2$ and the band has been centred around zero for convenience),
and $\theta(x)$ is the usual step function, and $u \equiv U g(\epsilon_F)$. Very often a different order parameter is favoured over $\phi(\omega_m)$, defined as 
$\Delta(\omega_m) \equiv \phi(\omega_m)/Z(\omega_m)$. Then Eqs.~(\ref{gb1},\ref{gb3}) are written as
\be
Z(i\omega_m)  =  1+ {\pi T_c \over \omega_m} \sum_{m^\prime = -\infty}^{+\infty}  \lambda(i\omega_m - i\omega_{m^\prime})
{\omega_{m^\prime}  \over \sqrt{\omega^2_{m^\prime} + \Delta^2(\omega_{m^\prime})}}
\label{gc1}
\ee
\be
Z(i\omega_m) \Delta(i\omega_m)  =  \pi T_c \sum_{m^\prime = -\infty}^{+\infty} 
\left[ \lambda(i\omega_m - i\omega_{m^\prime}) - u \ \theta ( {W \over 2} - |\omega_{m^\prime}| ) \right]
{ \Delta(i\omega_{m^\prime}) \over \sqrt{\omega^2_{m^\prime} + \Delta^2(\omega_{m^\prime})}}.
\label{gc3}
\ee

\ewt
These are the standard Eliashberg equations, written on the imaginary axis. These components of the self energy (refer back to
Eq.~(\ref{even_odd})) are all real functions. One further `simplification' is usually made before computations are performed. The Matsubara
summation is in principle infinite; in practice the summation with the electron-phonon kernel converges with relatively few terms, corresponding
to a frequency cutoff $\omega_c << W/2$. Yet in the term with the Coulomb interaction one is required to carry out a summation over many more terms,
corresponding to a frequency cutoff of $W/2$. Inspection of Eqs~(\ref{gc1},\ref{gc3}) shows that over this range ($\omega_c < \omega_m < W/2$), the gap 
function is a constant, i.e. $\Delta(\omega_m) \approx \Delta_\infty$, and $Z(\omega_m) \approx 1$. Making use of this allows one to sum this part analytically,
with the result that $U$ changes to an effective $U^\ast(\omega_c)$, where
\be
U^\ast(\omega_c) \equiv {U \over 1 + U {\rm ln}({W/2 \over \omega_c})},
\label{mustar}
\ee
and we have approximated digamma functions with their asymptotic logarithmic form, since it is assumed that both $W/2 >> T_c$ and $\omega_c >> T_c$. Now the
equations are
\bwt

\be
Z(i\omega_m)  =  1+ {\pi T_c \over \omega_m} \sum_{m^\prime = -\infty}^{+\infty}  \lambda(i\omega_m - i\omega_{m^\prime})
{\omega_{m^\prime}  \over \sqrt{\omega^2_{m^\prime} + \Delta^2(\omega_{m^\prime})}}
\label{gd1}
\ee
\be
Z(i\omega_m) \Delta(i\omega_m)  =  \pi T_c \sum_{m^\prime = -\infty}^{+\infty} 
\left[ \lambda(i\omega_m - i\omega_{m^\prime}) - u^\ast(\omega_c) \theta ( \omega_c - |\omega_{m^\prime}| ) \right]
{ \Delta(i\omega_{m^\prime}) \over \sqrt{\omega^2_{m^\prime} + \Delta^2(\omega_{m^\prime})}},
\label{gd3}
\ee

\ewt
where $u^\ast(\omega_c) \equiv g(\epsilon_F) U^\ast(\omega_c)$. One should note that $U^\ast(\omega_c)<~U$, physically corresponding to the fact
that retardation effects allow two electrons to exchange a phonon with one another while not being at the same place at the same time. This means
they do not feel the full direct Coulomb interaction with one another.

Thus far we have written the Eliashberg equations as functions of imaginary frequency. As we will see in the next subsection one can solve these equations as they are,
to determine many thermodynamic quantities of interest, in particular $T_c$. However, later we will extend these equations to the upper half-plane, and
in particular just above the real axis. This is required for the evaluation of dynamic quantities like the tunneling density of states and the optical conductivity.\cite{scalapino69,marsiglio97,marsiglio08} In anticipation of these results we note here that we use functions $Z(z)$ and $\phi(z)$ [and therefore $\Delta(z)$] with
the following properties\cite{ambegaokar64} as a function of complex frequency $z$
\bea
Z(z^\ast) = Z^\ast(z); &&\ \ \  Z(-z) = Z(z),\\
\phi(z^\ast) = \phi^\ast(z); &&\ \ \  \phi(-z) = \phi(z),\\
\Delta(z^\ast) = \Delta^\ast(z); &&\ \ \  \Delta(-z) = \Delta(z).
\label{gap_symmetries}
\eea

\subsection{Results on the imaginary axis: $T_c$}

To compute actual results for $T_c$, along with the gap function $\Delta(\omega_m)$ and the renormalization function $Z(\omega_m)$, we need to specify
$\alpha^2F(\nu)$ (now assumed to be isotropic) and $u^\ast(\omega_c)$. The latter quantity is very difficult to compute, and the former is more
tractable through Density Functional Theory. Historically it has been ``measured'' through tunnelling measurements.\cite{mcmillan69} We use
quotation marks around the word ``measured'' because in fact the current is measured while the spectral function is extracted through an inversion
process that requires theoretical input through the Eliashberg equations themselves.\cite{mcmillan69} We will simply adopt a model spectral
function given by
\be
\alpha^2F(\nu) = {\lambda_0 \nu_0 \over 2 \pi} \left[ { \epsilon \over (\nu - \nu_0)^2 + \epsilon^2} -  { \epsilon \over \nu_c^2 + \epsilon^2}  \right] \theta (\nu_c - |\nu - \nu_0|),
\label{model}
\ee
that is, a Lorentzian line shape cut off in such a way that the function goes smoothly to zero in the positive frequency domain. This Lorentzian has a centroid given
by $\nu_0$ and a half-width given by $\epsilon$. The cutoff frequency parameter $\nu_c$ makes the Lorentzian go to zero at frequency $\nu_0 + \nu_c$ and frequency
$\nu_0 - \nu_c$. For concreteness we will use a variety of values of $\nu_0$ with $\epsilon = 0$,  or $\epsilon \approx \nu_0/10$. The first choice results in a
$\delta$-function spectrum with weight such that the mass enhancement parameter, $\lambda$, defined by
\be
\lambda \equiv 2 \int_0^\infty d\nu {\alpha^2F(\nu) \over \nu} 
\label{lambda}
\ee
is simply given by $\lambda_0$. As $\epsilon$ increases $\lambda$ decreases from $\lambda_0$; however in what follows we will adjust $\lambda_0$
to keep $\lambda$ constant.\cite{remark1} Since the main focus of Eliashberg theory is the effect of retardation,
we will often set $u^\ast(\omega_c) = 0$, but we will nonetheless note how this quantity affects the gap function and $T_c$. 

Superconducting $T_c$ is determined by linearizing the gap equations, Eqs.~(\ref{gd1},\ref{gd3}) so that they become
\bwt

\be
Z(i\omega_m) = 1 + {\pi T_c \over \omega_m} \left\{ \lambda + 2 \sum_{n=1}^{m-1} \lambda(i\nu_n) \right\}.
\label{ge1}
\ee
\be
Z(i\omega_m) \Delta(i\omega_m)  =  \pi T_c \sum_{m^\prime = -\infty}^{+\infty} 
\left[ \lambda(i\omega_m - i\omega_{m^\prime}) - u^\ast(\omega_c) \theta ( \omega_c - |\omega_{m^\prime}| ) \right]
{ \Delta(i\omega_{m^\prime}) \over |\omega_{m^\prime}|}.
\label{ge3}
\ee

\ewt
The latter of these two equations is an eigenvalue equation and can
be solved as such. We use a power method that iterates the eigenvalue and eigenvector simultaneously by requiring that the gap function at 
the lowest Matsubara frequency,\cite{ewald} $\Delta(i\omega_1)$, remain at unity. This procedure tends to converge very quickly for stronger coupling, 
but requires more care for weaker coupling.\cite{marsiglio18}

% fig. 1 review
\begin{figure}[tp]
\begin{center}
\includegraphics[height=3.6in,width=4.1in]{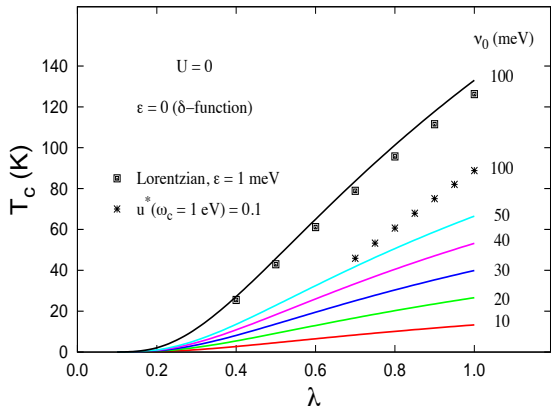}
\end{center}
\caption{The superconducting critical temperature, $T_c$ (K) vs. the dimensionless mass enhancement parameter, $\lambda$, for a variety of characteristic
phonon frequencies, as indicated. The solid curves are for the Einstein spectrum with $U = 0$. The square points indicate how $T_c$ changes (for $\nu_0 = 100$ meV
only) when a Lorentzian is used instead with $\epsilon = 10$ meV and $\nu_c = 80$ meV, according to Eq.~(\ref{model}).  For the same value of $\lambda$ there
is only a slight reduction in the value of $T_c$. The points marked with asterisks are again for the same broadened Lorentzian centred at $\nu_0 = 100$ meV, but
now with $u^\ast(\omega_c = 1$ eV. Note that $\nu_c$ is used as a practical cutoff for the phonon spectrum whereas $\omega_c$ is used for the Matsubara
cutoff for the direct Coulomb repulsion. See the discussion in the text for how plausible these parameters might or might not be.
}
\label{fig1}
\end{figure}
We begin with a standard plot of $T_c$ vs $\lambda$  in Fig.~1, for a simple $\delta$-function spectral function with frequency as shown.
There are scaling relations for $T_c$ with typical phonon frequency, but we choose to show the results explicitly in real units to make the 
result clear. The possibility of determining an expression for $T_c$ analytically has been discussed
extensively in the literature\cite{allen82} and will not be done here. The trends are clear; higher $T_c$ comes from higher values of $\lambda$ and from higher
values of the characteristic phonon frequency, which in the present case is provided by $\nu_0$. The width of the spectrum plays a minor role, and the Coulomb
repulsion suppresses $T_c$, as indicated by the marked points. It is well known that Eliashberg theory predicts that $T_c$ increases 
with both frequency and coupling strength as $T_c \approx \nu_0 \sqrt{\lambda}$ in the asymptotic limit.\cite{rainer73,allen75} 

\subsection{Validity of the Theory}

A perhaps more important question is the validity of the parameters 
used in the calculation. We have shown results up to a value of  $\lambda = 1$. Are higher values allowed? 
In particular, is it possible for a material to have a sizeable value of electron-phonon coupling while
maintaining a large phonon frequency? As discussed in the introduction, this question has been the subject of previous investigation,\cite{cohen72,moussa06,esterlis18}
although with only qualitative conclusions. The discovery of (very) high temperature superconductivity in the hydrides\cite{drozdov15,somayazulu19} under intense pressure
has spurred a reassessment of this type of analysis, since a part of the community believes that these superconductors are electron-phonon driven. The main evidence has
been an observed isotope shift.\cite{drozdov15} Moreover, the prediction of superconductivity in some of these compounds through density functional theory 
calculations\cite{duan14} adds plausibility to this explanation. However, very high characteristic phonon frequencies ($60$ - $120$ meV) and rather large electron-phonon
coupling values ($\lambda \approx 2$ or more) are required. The latter is well outside the range considered reasonable, {\it especially} given that the characteristic
phonon energy remains so high. Moreover, the superconductivity literature has unfortunately lapsed into simply accepting as ``standard'' or ``conventional'' a value
for the Coulomb pseudopotential $u^\ast = 0.1$, and, {\it especially} given the high value of phonon frequency, the anticipated reduction of the Coulomb interaction
through retardation will be much lower than previously thought, and the value of the direct Coulomb repulsion is undoubtedly higher when the phonon frequency is so high.

% fig. 2 review
\begin{figure}[tp]
\begin{center}
\includegraphics[height=2.4in,width=3.4in]{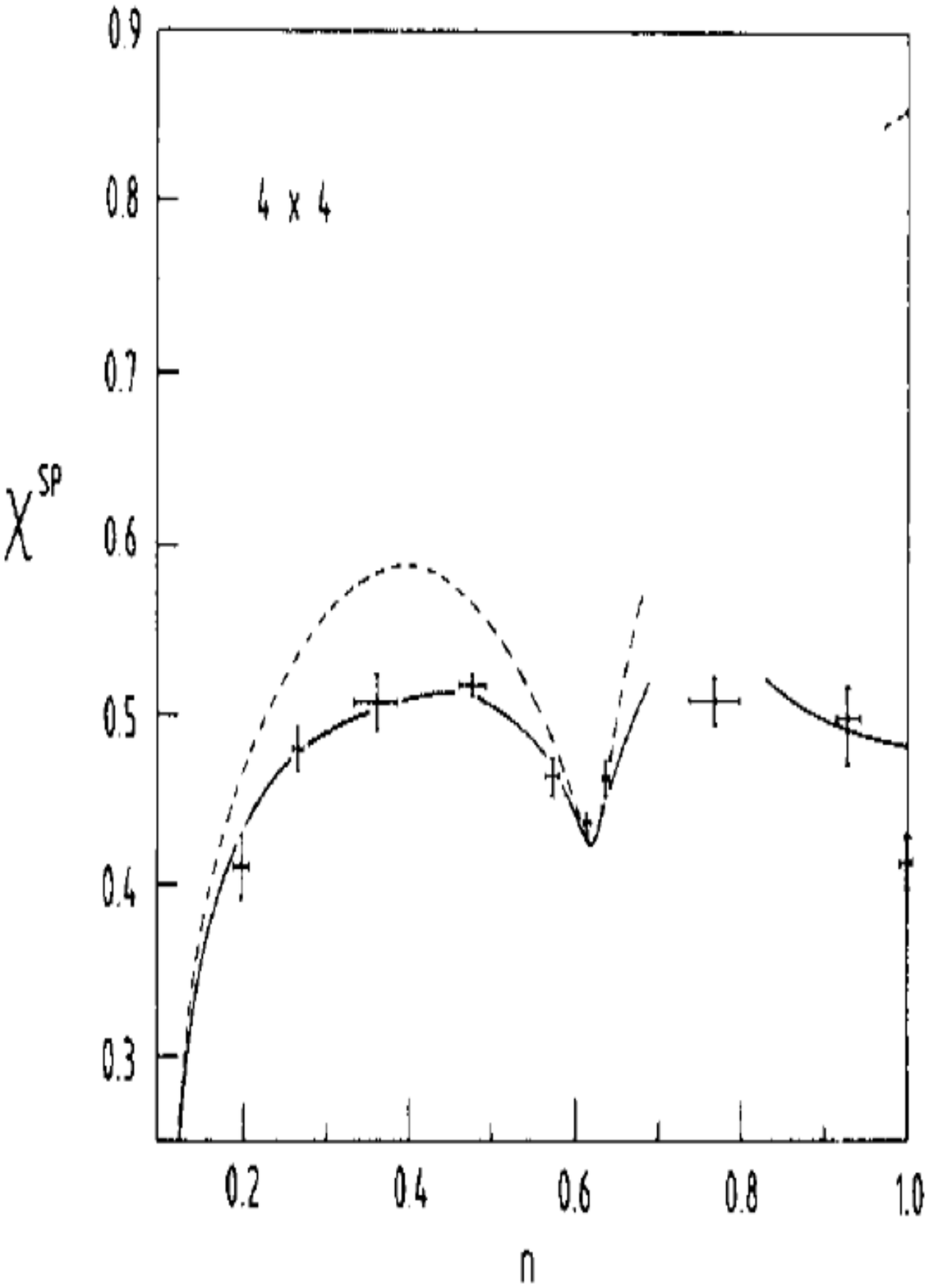}
\includegraphics[height=2.4in,width=3.4in]{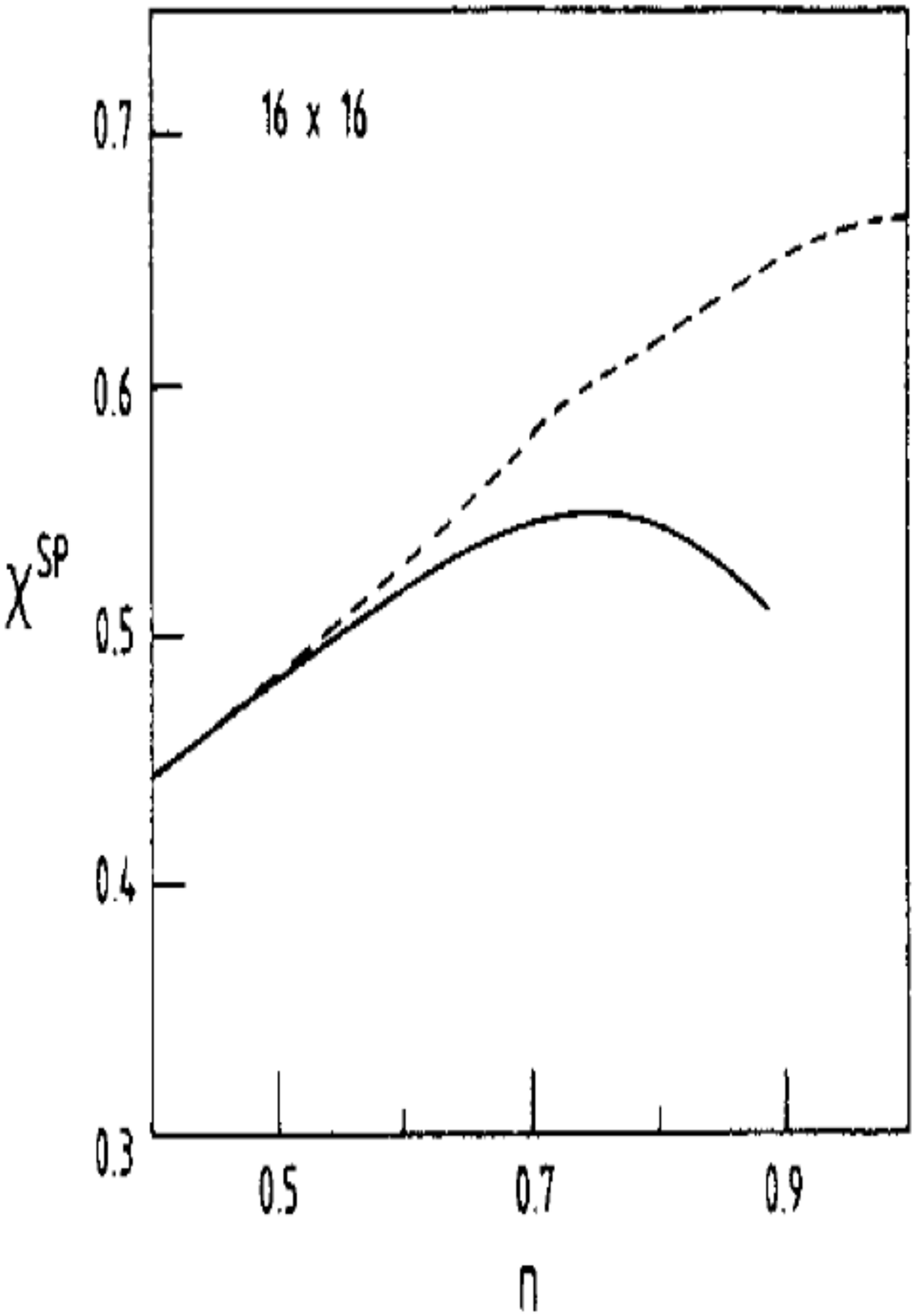}
\end{center}
\caption{The singlet pairing susceptibility vs. electron density for the Holstein model on a $4 \times 4$ lattice. See  Refs.~[\onlinecite{marsiglio90,marsiglio91}] for
pertinent definitions. Here a bare dimensionless coupling strength $\lambda_0 = 2$ and phonon (Einstein) frequency $\omega_E = 1t$ are used, and the susceptibility
is plotted for a temperature $T = t/6$, where $t$ is the nearest neighbour hopping parameter. In the topmost figure, QMC results are indicated with error bars. The solid
curves are the result for Migdal-Eliashberg theory with a renormalized phonon propagator and the dashed curves are the result for the unrenormalized calculations.
The renormalized calculations agree very well with the QMC results (both done for a $4 \times 4$ lattice), indicating that this (combined Migdal-Eliashberg
plus phonon self-energy in the bubble RPA approximation) result accurately captures the impact of CDW fluctuations on the pairing susceptibility. In the bottom
figure the renormalized (solid curve) and unrenormalized (dashed curves) are plotted for a larger system. The renormalized calculations stop at an electron density
close to $n \approx 0.9$ because a CDW instability occurs there. The unrenormalized calculations carry on to half-filling, because they are oblivious to the CDW
instability (and fluctuations). This result indicates that CDW fluctuations at densities less than $n = 0.9$ suppress pairing, and presumably $T_c$, even though
$\lambda^{\rm eff} \rightarrow \infty$. Reproduced
from Ref.~[\onlinecite{marsiglio91}]. 
}
\label{fig2}
\end{figure}

An additional direction of addressing this question comes from microscopic calculations involving Quantum Monte Carlo (QMC) and Exact Diagonalization (ED) techniques,
utilizing specific microscopic models. These methods provide controlled approximations and are therefore suitable for benchmarking more approximate 
theories like Eliashberg theory. By far the most work in this direction has been done on the Holstein model.\cite{holstein59} 
The Holstein model retains only the short-range (on-site) interaction between local (Einstein) oscillators and the electron charge density. Because it is 
a very local model it is more amenable to the exact or controlled methods developed over the past 40 years, and
therefore is a ``favourite'' for understanding the electron-phonon interaction, much like the Hubbard model\cite{hubbard63} is heavily used for the study of electron-electron interactions.  A short historical account of this activity is provided in the Appendix of Ref.~[\onlinecite{marsiglio08}]. 

Briefly, early Quantum Monte Carlo studies in one dimension\cite{hirsch82}
and two dimensions\cite{scalettar89,marsiglio90} established that charge-density-wave (CDW) correlations dominate at half-filling and close to half-filling. The critical
question is whether, sufficiently away from half-filling, where the susceptibility for superconductivity is stronger than that for CDW formation, do the ``remnant'' CDW
correlations {\it enhance} or {\it suppress} suppress superconducting $T_c$? In Ref.~[\onlinecite{marsiglio91}] the present author, based on a comparison of QMC
and Migdal-Eliashberg calculations on (very!) small lattices, argued that CDW fluctuations actually {\it suppress} superconducting $T_c$. In the so-called {\it renormalized}
Migdal-Eliashberg calculations a phonon self-energy was included; in this manner CDW fluctuations impacted the phonon propagator, 
resulting in softer phonons and an enhanced coupling constant. Comparisons with the QMC results served to benchmark the 
Eliashberg-like calculations. 

Figure 2, reproduced from Ref.~[\onlinecite{marsiglio91}],
illustrates that the so-called {\it renormalized} calculations agree with the QMC results. These calculations (solid curves) include phonon self-energy effects which are
essentially the CDW fluctuations.\cite{marsiglio90} In contrast, the {\it unrenormalized} calculations (dashed curves) are the standard Migdal-Eliashberg calculations that omit
phonon self-energy effects.  Figure~2(a) illustrates that the renormalized calculations are more accurate, and Fig.~2(b) shows that including CDW fluctuations
{\it suppresses} the pairing susceptibility, $\chi^{\rm SP}$. We understand these results to indicate that in the vicinity of a CDW instability, while the
effective coupling constant ($\lambda^{\rm eff}$ in Ref.~[\onlinecite{marsiglio90,marsiglio91}]) {\it increases}, superconducting $T_c$ {\it actually decreases}. 

More recently similar calculations have been performed\cite{esterlis18b} and other methodologies have been employed.\cite{bauer11} In the latter reference the role of retardation in reducing the
direct Coulomb interaction was also addressed; while they found qualitative agreement with the standard arguments, quantitative agreement was lacking, especially
for the expected large values of direct Coulomb repulsion. An older calculation with just two electrons,\cite{marsiglio95} based on Exact Diagonalization studies,
also found qualitative support for a retardation-related reduction in the direct Coulomb repulsion. It is worth mentioning that finding this insensitivity to increased
Coulomb repulsion, known as the ``pseudopotential effect,''\cite{bogoliubov59,morel62} has been looked for in QMC studies, but these have mostly been unsuccessful. They may still be there; part of the problem is that
QMC results become more difficult as the electronic and phonon energy scales differ from one another by a significant amount. Moreover, it may be that if larger
lattices and more realistic phonon frequencies (i.e. significantly less than the electron hopping parameter, $t$) are used, the result illustrated in Fig.~\ref{fig2} could
change qualitatively.

An additional concern has been raised about the electron-phonon coupling becoming too strong --- that of polaron collapse.\cite{alexandrov01} Exact 
studies in the thermodynamic limit\cite{bonca99} have established that a single electron, interacting with Einstein oscillators through the Holstein model, 
acquires an additional mass which is modest for $\lambda {{ \atop < }\atop {\approx \atop }} 1$, but becomes quickly (though smoothly!) very, very large beyond this point. This is
true independent of dimension,\cite{ku02,li12} and is especially acute when $\nu_0 << W$,\cite{li10}  where $W$ is the electronic bandwidth. So we have the intriguing
situation where the standard Migdal approximation (upon which Eliashberg theory is based) utilizes a Fermi sea of electrons strongly coupled to phonons, while
close examination of {\it just one} of these constituent quasiparticles (polarons) shows that it acquires a tremendous effective mass. Thus, the properties of the single
polaron, out of which a Fermi sea is constructed, appear to be incompatible with the properties of the electrons in that Fermi sea. 

For example, in Migdal theory, the effective
mass for electrons near the Fermi energy is $m^\ast/m_e \approx 1 + \lambda$, even for $\lambda \approx 2$ or more, whereas a single electron with this coupling would
have an effective mass many orders of magnitude higher. It is important to note that in the quantum treatment polarons never ``self-localize,'' essentially because
of Bloch's Theorem. However, with such large effective mass ratios, any impurities (including surfaces), would readily act as localization sites.

There are perhaps a few scenarios to work one's way out of this dilemma. First, as we have already mentioned, perhaps the Holstein model itself is pathological. For this
reason it is important to study other models. Unfortunately other models are more difficult to work with, but thus far the conclusions arrived at with the Holstein model
seem to hold for these other models as well. For example, in Ref.~[\onlinecite{peeters85}] a variational approach was used with the Fr\"ohlich Hamiltonian in the continuum
with acoustic phonons, and in Refs.~[\onlinecite{li11,chandler14}] the Barisi\'c-Labb\'e-Friedel/Su-Schrieffer-Heeger (BLF/SSH) model was examined with perturbation theory
and the adiabatic approximation. In either case more definitive results as achieved with the Holstein model are still lacking, although all indications are that these
models have strong polaronic tendencies as well.

A second scenario is that as one assembles a Fermi sea of polarons, they somehow become increasingly undressed, presumably due to some argument involving
Pauli blocking. There are no calculations that we are aware of, however, that provide a demonstration of this.\cite{hirsch95} Part of the reason may be psychological; the Migdal
approximation is more often called the Migdal Theorem, and so one may be inclined to take it for granted that this is what will happen when we assemble a
Fermi sea --- the ``theorem'' will be fulfilled. However, in my opinion this is more a belief than an established argument, as the Migdal approximation does not foresee
or account for polaron physics.

This discussion has been a long digression concerning the domain of applicability of Eliashberg theory, and clearly a lot more investigation is required on this question.
For now we return to the properties of the solutions to the Eliashberg equations.

\subsection{Results on the imaginary axis: in the superconducting state}

% fig. 3 review
\begin{figure}[tp]
\begin{center}
\includegraphics[height=3.2in,width=4.0in]{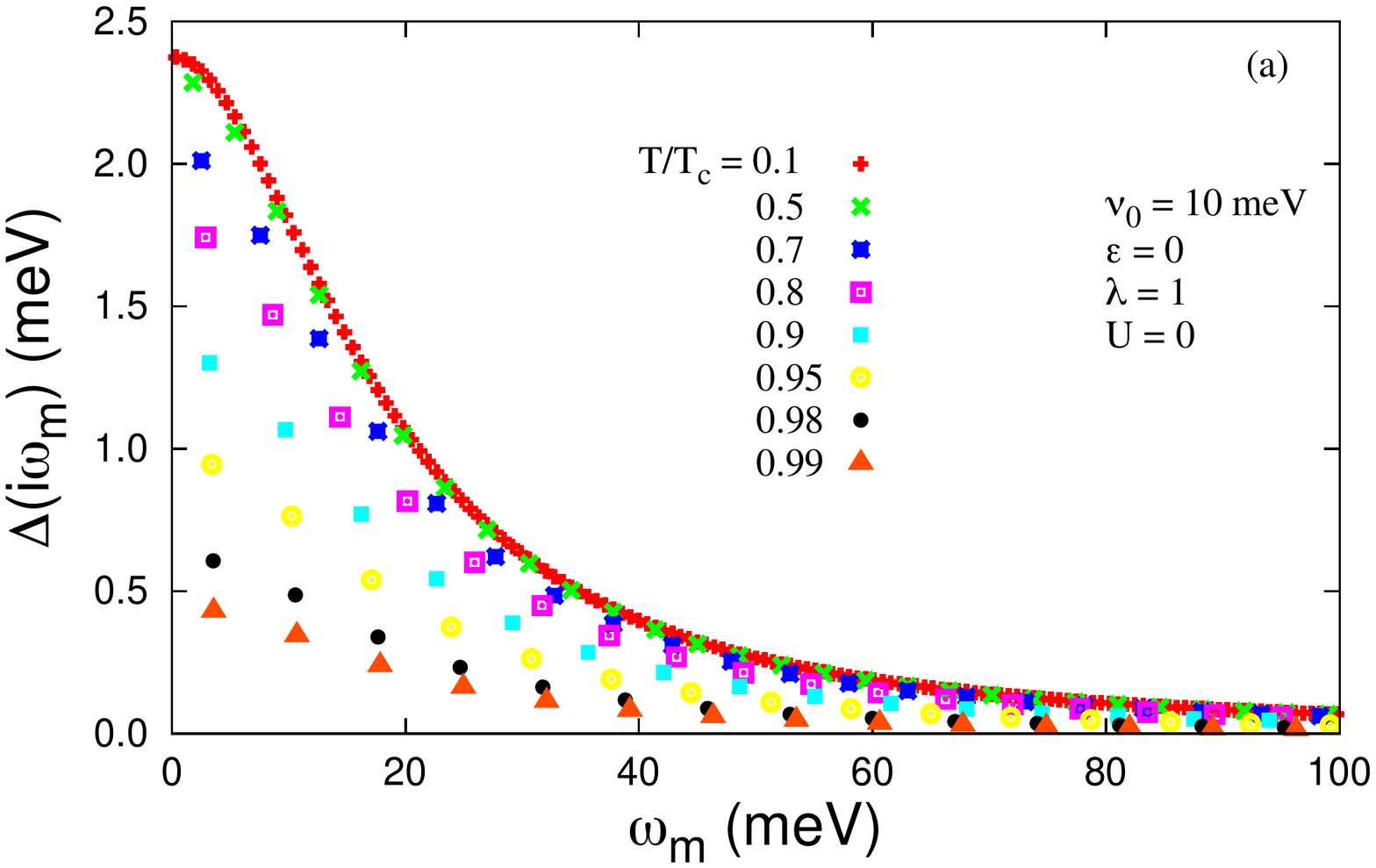}
\includegraphics[height=3.2in,width=4.0in]{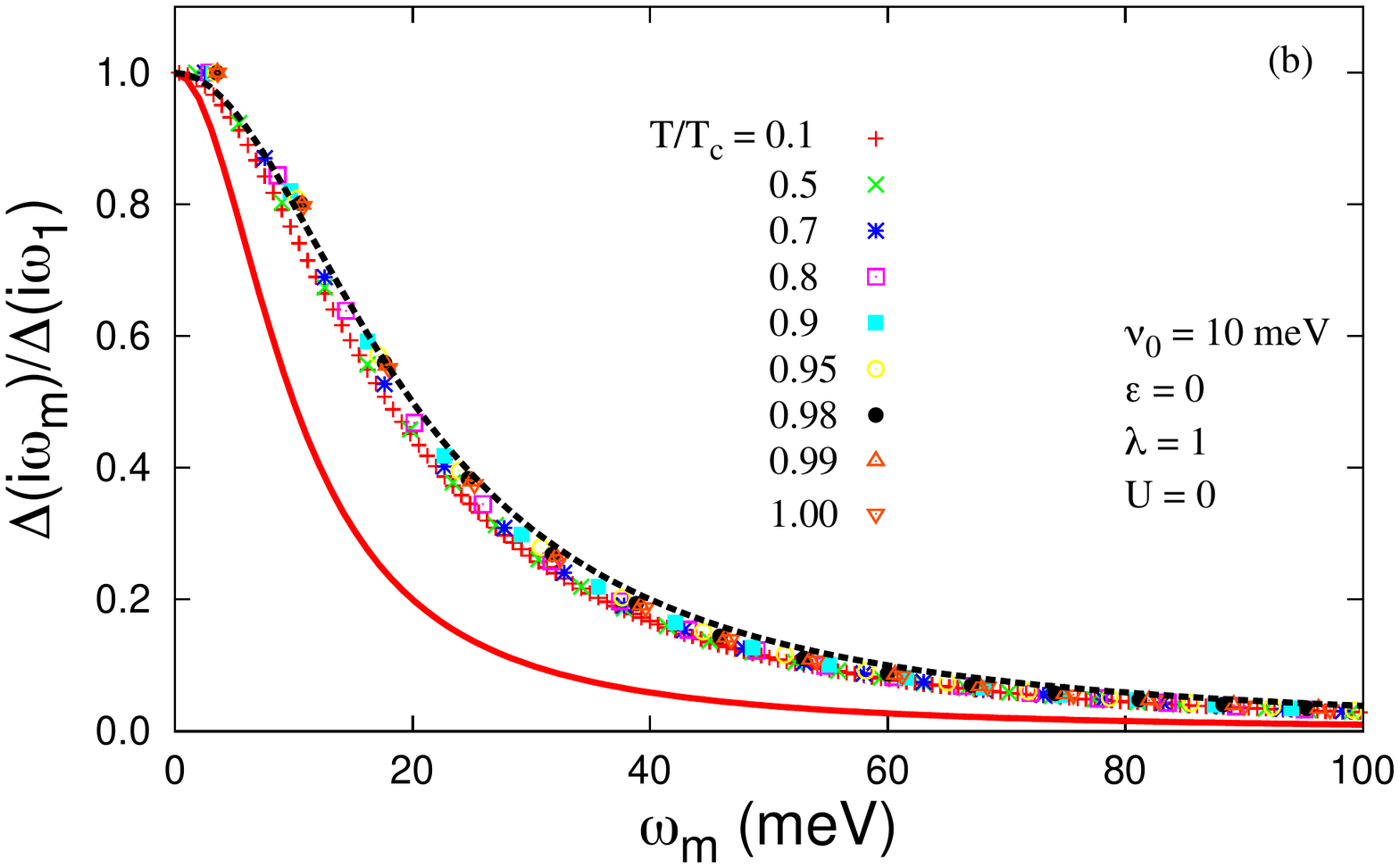}
\end{center}
\caption{(a) The gap function $\Delta(i\omega_m)$ vs. Matsubara frequency, $\omega_m$, for various temperatures. We have used a $\delta$-function phonon
spectrum ($\epsilon = 0$) with $\nu_0 = 10$ meV, electron-phonon coupling strength, $\lambda = 1$ and $U = 0$. For these 
parameters, $T_c = 13.3$ K. Clearly the gap function increases in amplitude with decreasing temperature, and below about $T/T_c = 0.5$ there is very little
change in the amplitude and in the frequency dependence. With a broadened phonon spectrum there would be only minor changes. With a nonzero $U$, the gap function
would have a negative asymptote as $\omega_m \rightarrow \infty$. In (b), to illustrate that there is very little change in the frequency dependence {\it at all temperatures} we
show the normalized gap function, $\Delta(i\omega_m)/\Delta(i\omega_1)$ vs. $\omega_m$. Now the results look very similar to one another, which makes
convergence from one temperature to the next relatively easy. Also shown is the weak coupling expectation\cite{marsiglio18} at $T_c$, 
$\Delta(i\omega_m)/\Delta(i\omega_1) = \omega_E^2/(\omega_E^2 + \omega^2_m)$, indicated with a solid red curve. This result clearly does not resemble the
data, since we the numerical results are for $\lambda = 1$. However, the slightly modified result,  $\Delta(i\omega_m)/\Delta(i\omega_1) = \omega_E^2/(\omega_E^2 + \omega^2_m/(1+\lambda)^2)$, shown as a dashed black curve, is a fairly good fit.}
\label{fig3}
\end{figure}

% fig. 4 review
\begin{figure}[tp]
\begin{center}
\includegraphics[height=3.2in,width=4.1in]{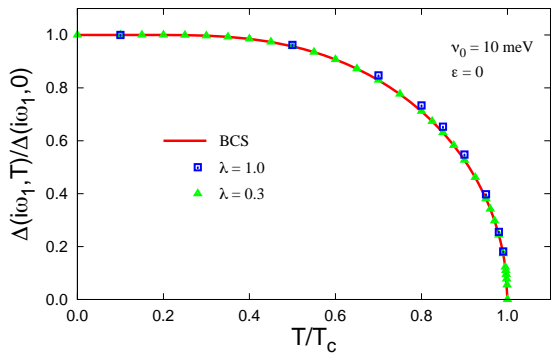}
\end{center}
\caption{The gap function at the first Matsubara frequency (serving as an order parameter), normalized to the zero temperature gap function at the first Matsubara 
frequency, vs. reduced temperature $T/T_c$. The blue squares are the results at a few temperatures for calculations using a phonon $\delta$-function spectrum with
$\nu_0 = 10$ meV, $\lambda = 1$, and $U=0$. Shown for comparison is the weak coupling BCS result (red curve). The deviations are very slight. Also shown for comparison
are the Eliashberg results for the same phonon spectrum but with $\lambda = 0.3$ (green triangles). These results fall right on the BCS weak coupling result.
}
\label{fig4}
\end{figure}
Returning to Eqs.~(\ref{gd1},\ref{gd3}), or their linearized counterparts, Eqs.~(\ref{ge1},\ref{ge3}), once $T_c$ is determined then the gap function can be determined
both at $T_c$ and below $T_c$. The gap function is a generalized (frequency-dependent) order parameter. It will grow continuously from zero at $T_c$ to its
full value at zero temperature, but it depends on frequency. In Fig.~\ref{fig3}(a) we show the gap function as a function of Matsubara frequency for a variety of temperatures,
for $\lambda = 1$ and $\nu_0 = 10$ meV. Note that these functions are defined on a discrete set of points (the Fermion Matsubara frequencies) that become more
closely spaced as the temperature is lowered.
For temperatures close to $T_c$ the gap function diminishes gradually to zero at all frequencies, while at the lowest temperature the gap function attains a maximum. In 
Fig.~\ref{fig3}(b) we plot the normalized values, $\Delta(i\omega_m)/\Delta(i\omega_1)$ vs. Matsubara frequency, and it is clear that they differ from one another by
very little. Returning to Fig.~\ref{fig3}(a), the lowest frequency function value can be thought of as an order parameter. In Fig.~\ref{fig4} we show the lowest frequency
gap function value, $\Delta(i\omega_1)$, now normalized to the value at $T=0$ vs. reduced temperature, $T/T_c$. These are shown as blue squares, for about 9
temperatures. Also shown is the BCS weak coupling result, given as a red curve. One can see that the differences are small. We have also a plotted many more points
(green asterisks) for a weaker coupling, $\lambda = 0.3$ (same phonon frequency), which fall exactly on the BCS curve. The main point is that deviations from the
weak coupling BCS result are minor. For much stronger coupling than given here deviations are similarly very small, and experiment confirms this to be the 
case.\cite{gasparovic66}

Many measurable properties of the superconducting state can be calculated from the imaginary axis solutions to the gap function. The renormalization
function, $Z(i\omega_m)$, is also required but this does not change by very much in the superconducting state. Examples of measurable properties
include all thermodynamic
quantities like the specific heat, and various critical fields. Systematic changes with coupling strength, as measured by $\lambda$, or alternatively the
ratio of the critical temperature to a particular phonon frequency moment, $T_c/\omega_{\rm ln}$, have been reviewed elsewhere,\cite{carbotte90,marsiglio08}
and will not be repeated here.

\subsection{Results on the real axis}

For any dynamical property (tunneling current, optical response, dynamical penetration depth, etc.), the relevant Green function (and therefore self-energy)
is required as a function of real frequency. More precisely, for the retarded Green function it is needed at $\omega + i\delta$, i.e. infinitesimally above the real axis.
In the original literature\cite{eliashberg60a,eliashberg60b,morel62,schrieffer63,scalapino66} the spectral representation was introduced to replace
Matsubara sums with real frequency integrals, and these equations were then solved, either analytically (with approximations) or numerically. This was a difficult
task (especially with the computers available at that time), and eventually the procedure was adopted that first required a solution on the imaginary axis and then
analytic continuation ($i\omega_m \rightarrow \omega + i\delta$) through some approximate process. 
For this type of analytic function, the method of Pad\'e approximants was used,\cite{vidberg77}
although the degree of precision needed for the gap function on the imaginary axis was very stringent ($10^{-12}$ for relative errors) in order
to achieve accurate results on the real axis. 

An appreciation for the information imbedded in imaginary axis solutions can be attained by considering
the simple example of $g(i\omega_m) = {\rm sech}(\omega_m/\nu_0)$, a very smooth function without structure, and monotonically decreasing with
frequency on the positive imaginary axis. The analytic continuation can be easily done analytically; it is $g(\omega + i\delta) = {\rm sec}[(\omega + i\delta)/\nu_0]$.
This function, in contrast to its imaginary axis counterpart, is riddled with divergences and discontinuities. And yet, in principle at least, this information
is embedded in the (smooth) results on the imaginary axis. In practice, the information is contained in the $10^{\rm th}$ significant digit and beyond.

An alternative, numerically exact procedure was developed in the late 1980's.\cite{marsiglio88} Here we simply write down the resulting expressions, the derivation
of which is available in Ref.~[\onlinecite{marsiglio88}]. They are
\bwt

\begin{eqnarray}
\phi(\omega + i \delta)  &=& \pi T\sum_{m=-\infty}^{\infty} \bigl[
\lambda(\omega-i\omega_m)-u^*(\omega_c)\theta(\omega_c-|\omega_m|)
\bigr] 
%{\phi({i\omega_m}) \over \sqrt{\omega_m^2 Z^2({i\omega_m})
%+ \phi^2({i\omega_m})} }
{\Delta({i\omega_m}) \over \sqrt{\omega_m^2 + \Delta^2({i\omega_m})} }
\nonumber \\
 +  i\pi\int_0^{\infty}d\nu\,\alpha^2F(\nu) && \Biggl\{\bigl[N(\nu)+
f(\nu-\omega)
\bigr] {\phi(\omega-\nu + i \delta) \over \sqrt{(\omega-\nu)^2
Z^2(\omega-\nu + i \delta) -
\phi^2(\omega-\nu + i \delta)}}
\nonumber \\
&& + \bigl[N(\nu)+f(\nu+\omega)\bigr] {\phi(\omega+\nu + i \delta)
\over \sqrt{(\omega+\nu)^2 Z^2(\omega+\nu + i \delta) -
\phi^2(\omega+\nu + i \delta)}}\Biggr\}\;,\nonumber \\
\label{gf1}
\end{eqnarray}
and
\begin{eqnarray}
Z(\omega + i\delta)  &=&  1 + {i\pi T \over \omega}\sum_{m=-\infty}^{\infty} \lambda(\omega-i\omega_m) 
%{\omega_m Z({i\omega_m}) \over 
%\sqrt{\omega_m^2 Z^2({i\omega_m}) + \phi^2({i\omega_m})} } \nonumber \\
{\omega_m  \over 
\sqrt{\omega_m^2  + \Delta^2({i\omega_m})} } \nonumber \\
+ {i\pi\over \omega} \int_0^{\infty}d\nu\,\alpha^2F(\nu) && \Biggl\{\bigl[N(\nu)+f(\nu-\omega)
\bigr]  {(\omega-\nu)Z(\omega-\nu + i \delta) \over \sqrt{(\omega-\nu)^2 Z^2(\omega-\nu + i \delta) - \phi^2(\omega-\nu+i
\delta)}} \nonumber \\ & & +\bigl[N(\nu)+f(\nu+\omega)\bigr]
{(\omega+\nu)Z(\omega+\nu + i \delta) \over
\sqrt{(\omega+\nu)^2 Z^2(\omega+\nu + i \delta) -
\phi^2(\omega+\nu + i \delta)}}\Biggr\}\;, 
\label{gf3}
\end{eqnarray}

\ewt
and of course $\Delta(\omega + i\delta) \equiv \phi(\omega + i\delta)/Z(\omega + i\delta)$. Here $f(\omega) \equiv 1/({\rm exp}(\beta \omega) + 1)$ and
$N(\nu) \equiv 1/({\rm exp}(\beta \nu) - 1)$ are the Fermi and Bose functions respectively.
Note that in cases where the square--root is complex, the branch with positive imaginary part is to be chosen. The reason for this can be traced back
to Eq.~(\ref{ga3}) [or Eq.~(\ref{ga1})] where the integration over $\epsilon_{{\bf k^\prime}}$ (with the assumptions made there) requires that the pole
(given the same square--root that appears here) be {\it above} the real axis.

It can easily be verified that substituting $\omega + i\delta \rightarrow i\omega_n$ instantly recovers the imaginary axis equations (all the Fermi and
Bose factors cancel to give zero contributions beyond the initial terms that require Matsubara summations). Clearly the inverse is not true --- replacing the
Matsubara frequency $i\omega_m$ where it appears in Eqs.~(\ref{gd1},\ref{gd3}) produces the first lines in Eqs.~(\ref{gf1},\ref{gf3}) involving Matsubara sums, but
leaves out the remaining two lines in each case. The strategy for solving these equations is straightforward; the imaginary axis equations [Eqs.~(\ref{gd1},\ref{gd3}) ]
are first solved self-consistently. These are then used in Eqs.~(\ref{gf1},\ref{gf3}) and these equations are iterated to convergence. The presence of the first lines in these
equations provides a ``driving term'' that makes the iteration process quite rapid. For example, performing the entire operation (solution of imaginary axis and
real axis equations) for a given temperature takes about a tenth of a second on a laptop.

Moreover, $T=0$ is a special case, as is clear from these equations. In fact, this was recognized a long time ago,\cite{karakozov75} where they established
the following low frequency behaviour at $T=0$,
\begin{eqnarray}
 { {\rm Re} \Delta(\omega + i\delta) \, = \, c ,\atop {\rm Im} \Delta(\omega + i\delta) \,
= \, 0,} \ \ \ \ \ \ \ \ \ T = 0
\nonumber \\
{ {\rm Re} Z(\omega + i\delta) \, = \, d , \atop {\rm Im} Z(\omega + i\delta) \, = \, 0.} \phantom{ \ \ \ \ \ \ \ \ \ T = 0}
\label{zerot_freq}
\end{eqnarray}
where $c$ and $d$ are constants. In contrast, the behaviour at any non-zero temperature is
\begin{eqnarray}
{ {\rm Re} \Delta(\omega + i\delta) \, \propto \, \phantom{aa}  \omega^2, \atop {\rm Im}
\Delta(\omega + i\delta) \, \propto \, \phantom{aa} \omega,}  \ \ \ \ \ \ \ \ \ T > 0
\nonumber \\
\nonumber \\
{ {\rm Re} Z(\omega + i\delta) \, = \, d(T), \atop {\rm Im} Z(\omega + i\delta) \,
\propto \, 1/\omega .}  \phantom{ \ \ \ \ \ \ \ \ \ T > 0}
\label{abovezerot_freq}
\end{eqnarray}
For conventional parameter choices this distinction has very little consequence, as the differences are barely discernible. For example, the expression
of the imaginary part of $Z(\omega + i\delta)$ in the normal state is given by
\be
{\rm Im}Z(\omega + i\delta) = 2 \pi \lambda {\nu_0 \over \omega}[N(\nu_0) + f(\nu_0)]
\label{normal_gamma}
\ee
for a $\delta$-function spectrum [$\epsilon \rightarrow 0$ in Eq.~(\ref{model})] with strength $\lambda$ and central frequency $\nu_0$. For $\nu_0 = 10$ meV
and $\lambda = 1$ then Fig.~\ref{fig1} indicates $T_c \approx 10$ K, and at $T/T_c = 0.1$, the exponent in the Bose and Fermi functions makes 
this part $\approx 10^{-50}$. Thus, true to Eq.~(\ref{abovezerot_freq}) the limiting behaviour is $\propto \nu_0/\omega$. However, to see this requires
$\omega/\nu_0 < 10^{-45}$, making it unobservable, and indistinguishable from the $T=0$ case. 
% fig. 5 review 
 \bwt
 
 \begin{figure}[htb]
%  \begin{figure}[tp]
\centering
  \begin{tabular}{@{}cc@{}}
%  \includegraphics[height=2.86in,width=2.3in,angle=-90]{fig5arev.eps} &
%\includegraphics[height=2.86in,width=2.3in,angle=-90]{fig5brev.eps} \\
%\includegraphics[height=2.86in,width=2.3in,angle=-90]{fig5crev.eps} &
%\includegraphics[height=2.86in,width=2.3in,angle=-90]{fig5drev.eps}\\
%    \multicolumn{2}{c}{\includegraphics[height=3.9in,width=2.7in,angle=-90]{fig5erev.eps}} 
      \includegraphics[height=2.4in,width=2.7in]{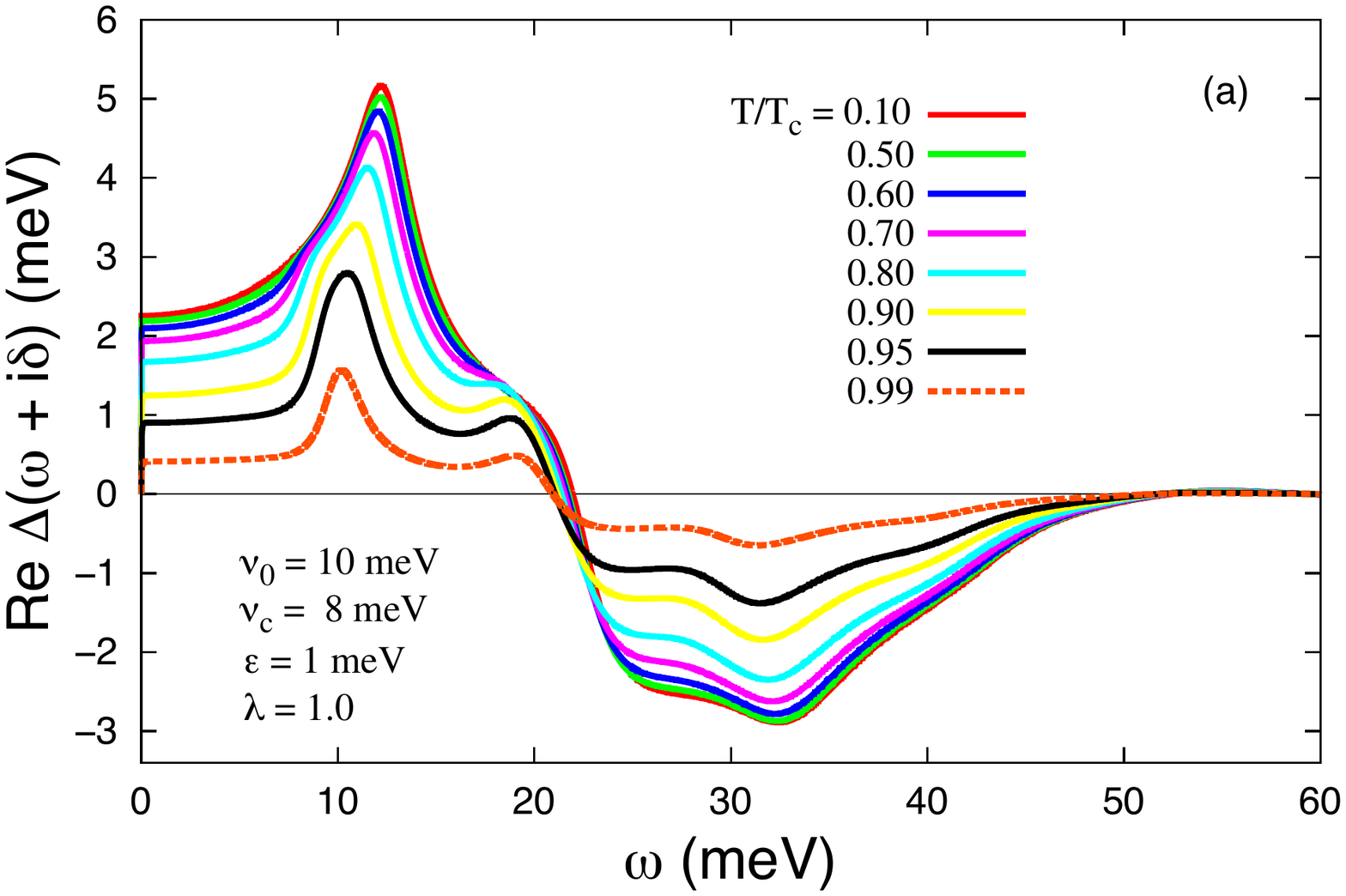} &
\includegraphics[height=2.4in,width=2.7in]{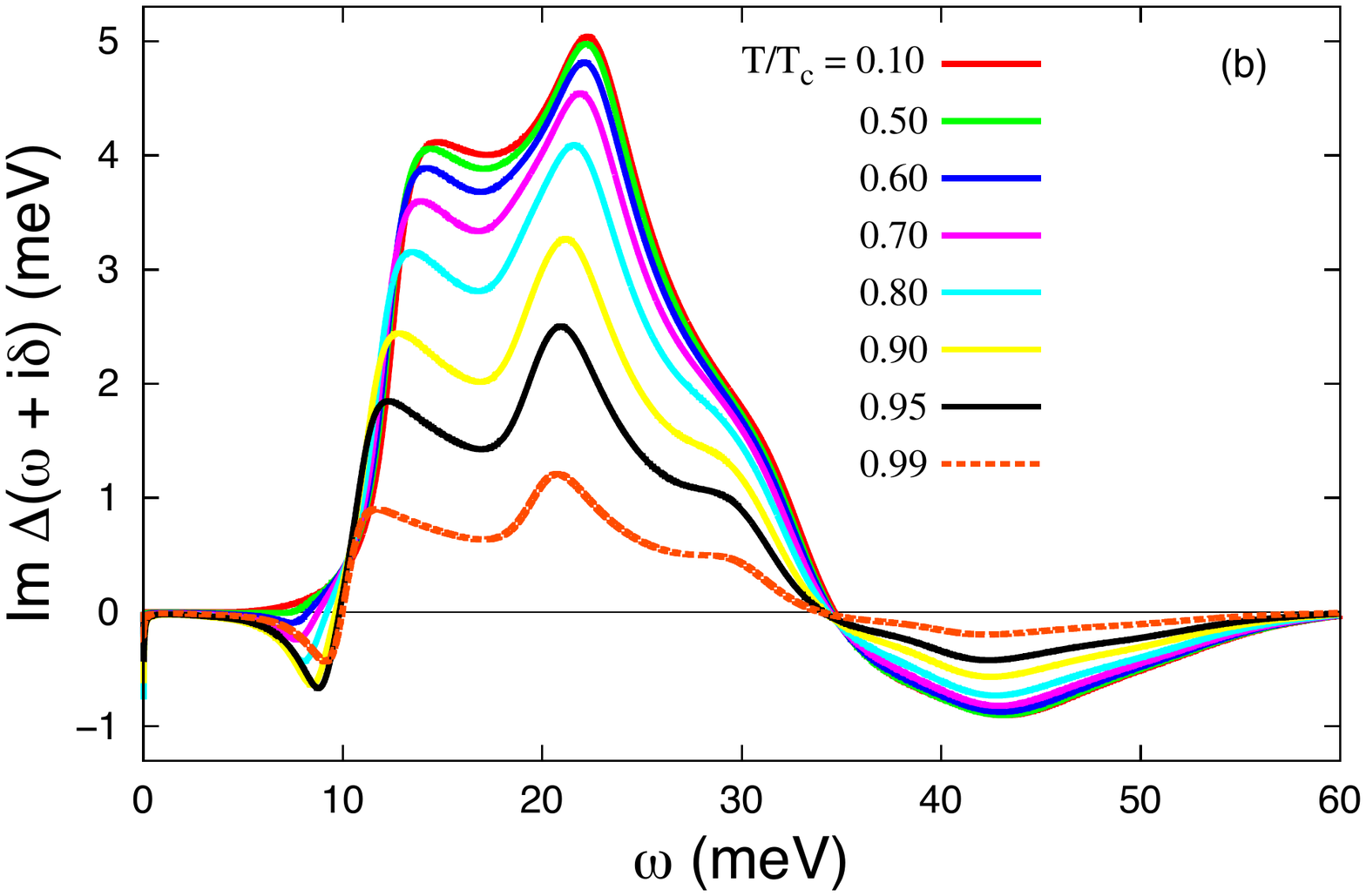} \\
\includegraphics[height=2.4in,width=2.7in]{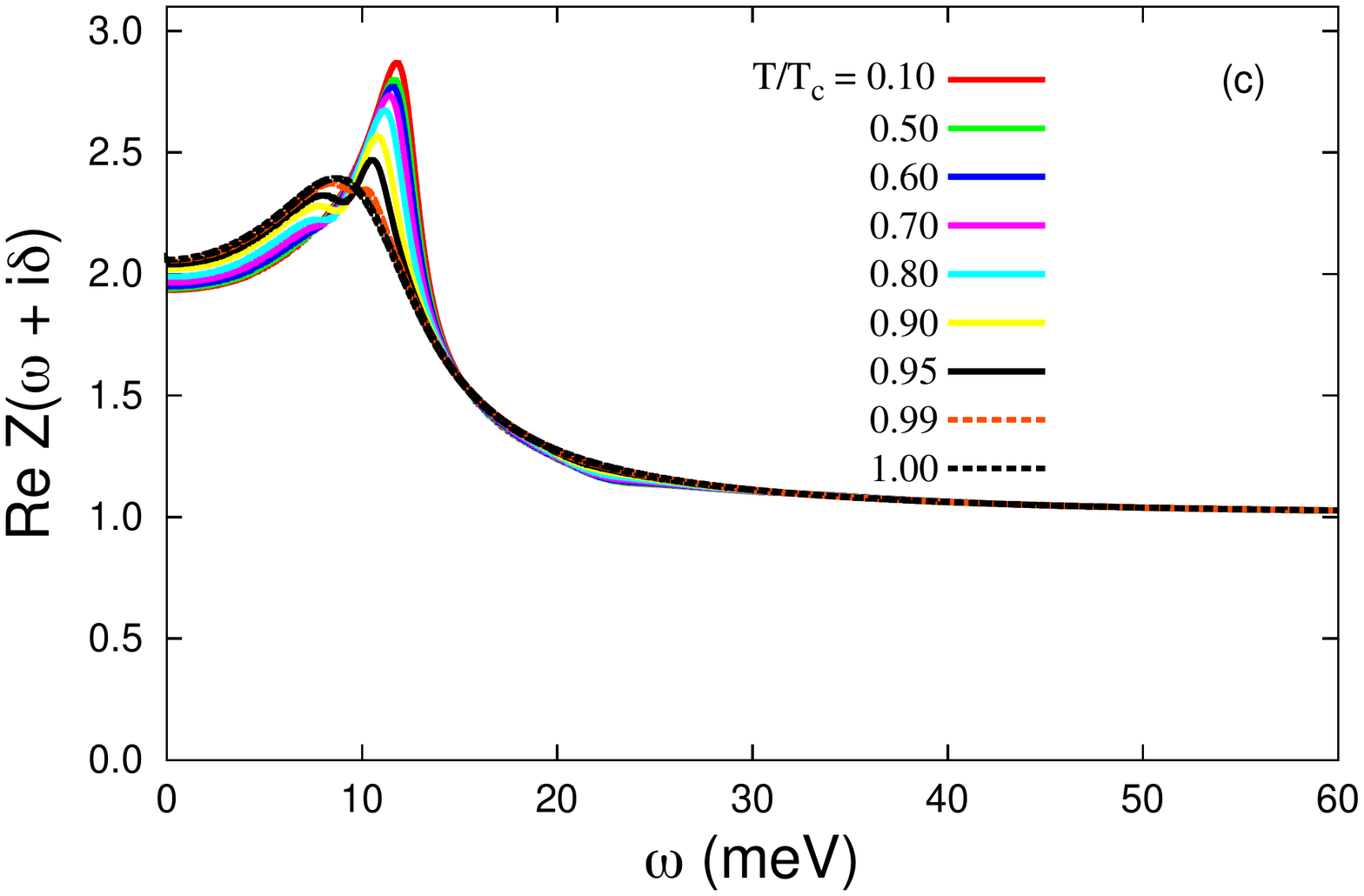} &
\includegraphics[height=2.4in,width=2.7in]{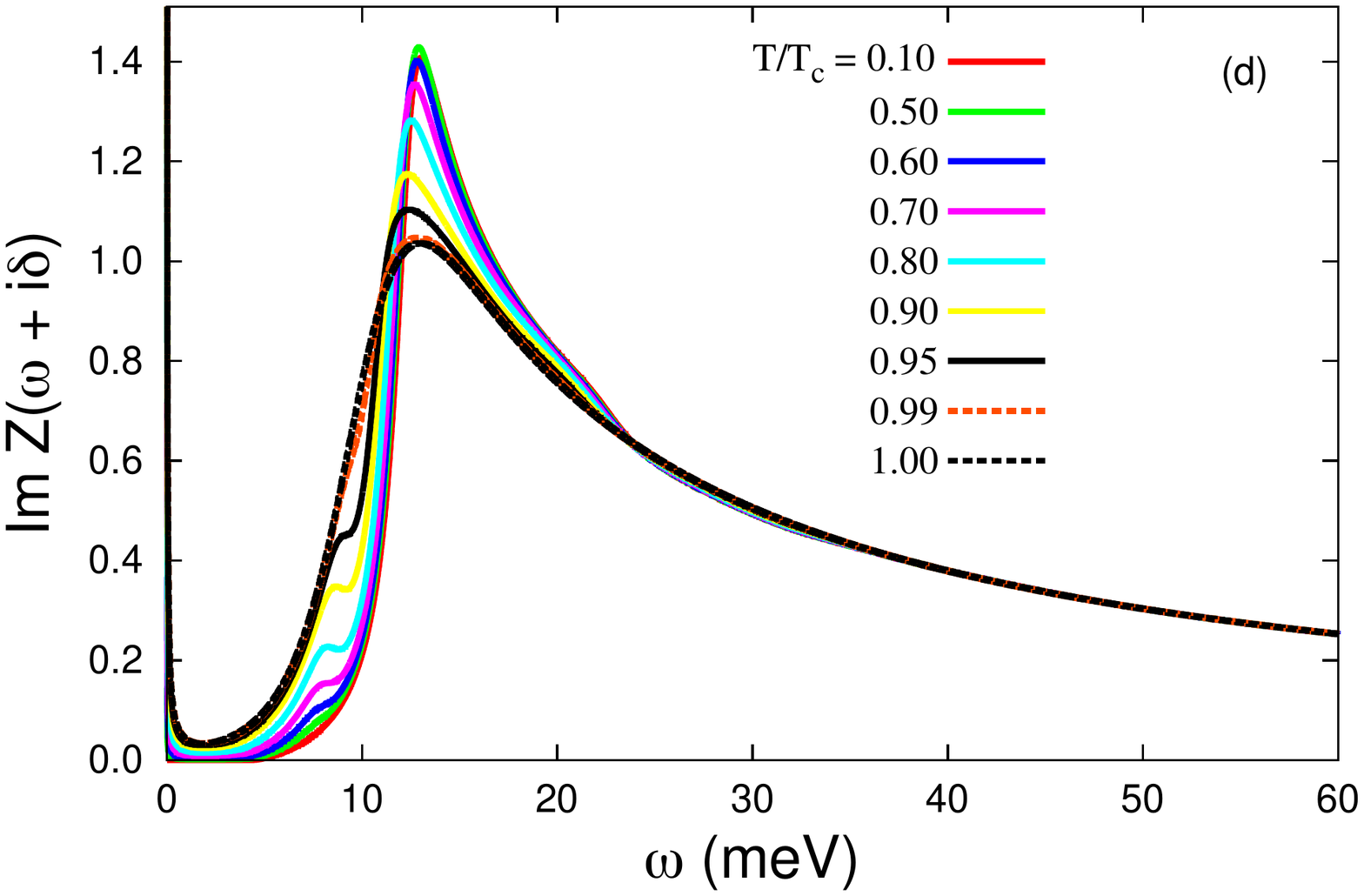}\\
    \multicolumn{2}{c}{\includegraphics[height=2.4in,width=2.7in]{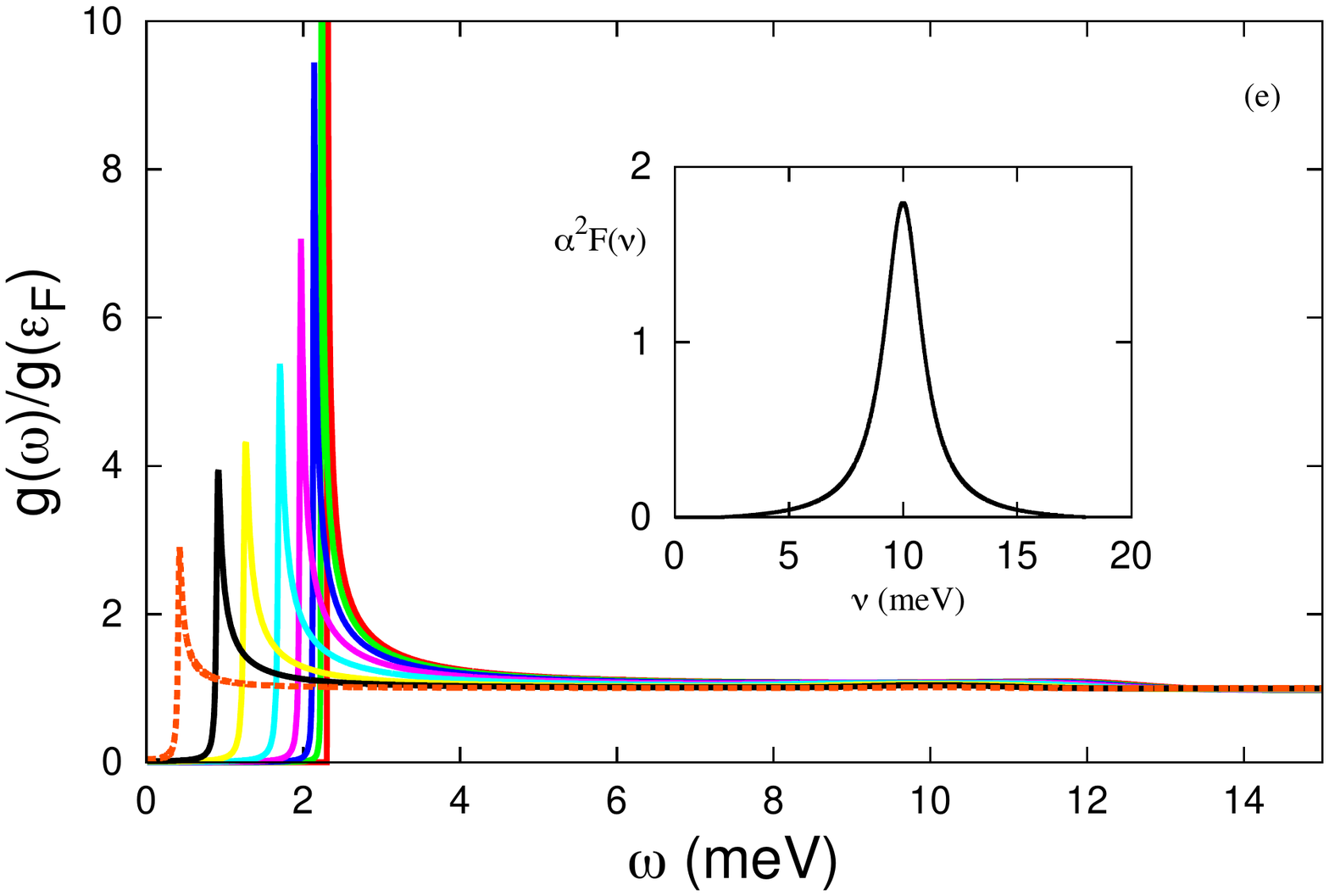}} 
  \end{tabular}
\caption{Frequency dependence of (a) ${\rm Re}\ \Delta(\omega+ i \delta)$, (b) ${\rm Im}\ \Delta(\omega+ i \delta)$, (c) ${\rm Re}\ Z(\omega+ i \delta)$, 
(d) ${\rm Im}\ Z(\omega+ i \delta)$, and (e) $g(\omega)/g(\epsilon_F)$, for various temperatures in the
superconducting state. Features are discussed in the text. For the spectral function we have used Eq.~(\ref{model}) with $\lambda = 1$, $\nu_0 = 10$ meV,
and $\epsilon = 1$ meV. It is displayed in the inset of (e) and was used for all these figures. The colour coding in (e) is the same as the others. We used $U=0$. 
}
\label{fig5}
\end{figure}

\ewt
While this expression is for the normal state, the corresponding one of the superconducting state is even more severe, due to the development of the
superconducting order parameter, which gives rise to a gap in the spectrum. Again, given the first two lines in Eq.~(\ref{abovezerot_freq}) there is
technically {\it no gap}, but in practice for reasons like we have just indicated, the practical results more closely follow the behaviour indicated in
Eq.~(\ref{zerot_freq}). Where the finite temperature result becomes pronounced and noticeably different than the zero temperature behaviour is in the strong coupling limit,\cite{marsiglio91b} but in this case the parameters are not realistic and undoubtedly beyond the domain of validity of the theory.

Returning to $T=0$, for cases like the present where the phonon spectrum has a gap there is a special simplification. Basically, no iteration is required --- the low frequency
gap and renormalization functions come entirely from the first lines of Eqs.~(\ref{gf1},\ref{gf3}), and these can be constructed explicitly from the imaginary axis results. For
a $\delta$-function phonon spectrum, however, one has to be careful to convert the Matsubara summation to an actual integral, as a discontinuity will occur at the
phonon frequency (at non-zero temperature this discontinuity is broadened into a gradual drop). Once the low frequency gap and renormalization 
functions are so constructed, higher frequency values require the Matsubara sum {\it and}
real axis values of these functions at {\it lower frequencies only}. Eventually the entire functional forms are so constructed, without the need for iteration. As we will
see, low temperature results converge quite rapidly to the zero temperature result, so this non-iterative method can be used as an alternative, for the lowest temperatures. Nonetheless, we will proceed with fully converged (iterative) finite temperature results, since these require such few iterations anyways.

To show real axis results, we utilize a phonon spectrum as in Eq.~(\ref{model}) with $\nu_0 = 10$ meV, $\nu_c = 8$ meV, and $\epsilon = 1$ eV, and $\lambda = 1$. 
Results with a $\delta$-function spectrum tend to have a series of singularities, that are anyways artifacts of the singular spectrum, so we prefer to show results
corresponding to this model spectrum. A series 
of such plots was also shown in Ref.~[\onlinecite{marsiglio08}] for the tunneling-derived Pb spectrum, with a much larger value of $\lambda$, and many more
such results have been shown in the literature, often using this same method.\cite{carbotte19} In Fig.~\ref{fig5} we show (a) the real part of the gap function, 
(b) the imaginary part of the gap function, (c) the real part of the renormalization function, (d) the imaginary part of the renormalization function, and (e) the
tunneling density of states, 
\be
{g(\omega) \over g(\epsilon_F)} = {\rm Re} { \omega \over \sqrt{\omega^2 - \Delta^2(\omega + i \delta)}},
\label{tunnel}
\ee
which is measurable in single-particle tunneling experiments. The first observation, difficult to make with just these results, is that an image of the
$\alpha^2F(\nu)$ spectrum is contained in both the real and imaginary parts of the gap function. Here it is the peak structure clearly evident in (a)
centred around 10 meV {\it for the highest temperature shown.} As the temperature decreases this peak shifts to higher frequency, roughly by an amount
equal to the value of the gap function at low frequency (about $2$ meV in the present case). Experimentation with different spectral functions makes this
observation more self-evident. See, for example, the distinctive spectrum for Pb in Fig.~4.35(a) of Ref.~[\onlinecite{marsiglio08}]. 

Both functions in (a) and (b) go to zero as the critical temperature is approached from below. As discussed earlier, they both go to zero at zero frequency at all
temperatures shown, according to Eqs.~(\ref{abovezerot_freq}), although one cannot see this on the scale shown. Even for the highest temperatures shown this behaviour
can barely be seen, but becomes evident when one expands the low frequency scale. For the lowest temperatures shown even expanding the frequency scale by
a few orders of magnitude is not enough to reveal the low-frequency behaviour indicated by Eqs.~(\ref{abovezerot_freq}). For this reason one cannot use 
$\Delta(\omega + i\delta)$ as an order parameter at any frequency; either one has to revert to $\phi(\omega + i\delta)$ at zero frequency, or one can use
the imaginary axis result for $\Delta(i\omega_m)$, as we did in Fig.~\ref{fig4}. Also note that these functions approach zero at high frequency. If a Coulomb repulsion
is included then the real part of $\Delta(\omega + i\delta)$ approaches a negative constant at high frequency.

In contrast the real and imaginary parts of $Z(\omega + i\delta)$ plotted in (c) and (d) have changed very little in the superconducting state, and remain non-zero
at the superconducting critical temperature, as indicated by the black curve. An image of $\alpha^2F(\nu)$ is present in this function as well, particularly in the
imaginary part (see also Fig.~4.35 (c) and (d) in Ref.~[\onlinecite{marsiglio08}]). Finally, the tunneling density of states is shown in Fig.~\ref{fig5}(e), and reveals a
``gap'' that opens from zero at $T=T_c$ rather quickly and then saturates to a low temperature value as indicated. In fact a plot of this ``gap'' vs. temperature
would follow the result displayed in Fig.~\ref{fig4} very closely. However, as first pointed out by Karakozov et al.\cite{karakozov75} there is no ``gap'' (hence the
parentheses) and in fact this is evident in Fig.~\ref{fig5}(e), where there is a noticeable rounding of the curves at frequencies below the sharp peak at almost all
temperatures shown. The peak is a remnant of the square-root singularity known from BCS theory, which is evident from Eq.~(\ref{tunnel}) if a constant gap
function is used, $\Delta(\omega+i\delta) = \Delta_0$. In fact Eliashberg theory predicts smearing of this singularity simply due to the presence of imaginary
components of all the functions involved in Eqs.~(\ref{abovezerot_freq}) for all finite temperatures. It is also worth pointing out that the BCS limit of
Eliashberg theory is {\it not} achieved by setting the gap function to a constant, $\Delta(\omega+i\delta) \rightarrow \Delta_0$, but in fact the gap function
is a decaying function of frequency in this limit.\cite{karakozov75} This frequency dependence and its implications for the weak coupling limit has been 
further explored in Refs.~[\onlinecite{wang13,marsiglio18,mirabi19}].

Finally, though not so evident in Fig.~\ref{fig5}(e), there are ``ripples'' in the density of states beyond the ``gap'' region, caused by the coupling of electrons
to phonons. The presence of these ripples in experiments (see, in particular, Refs.~[\onlinecite{rowell63,mcmillan65,mcmillan69}]) is perhaps the strongest
evidence of the validity of Eliashberg theory. In fact the most intense scrutiny has been superconducting Pb, where the electron-phonon coupling is particularly
strong, $\lambda \approx 1.5$, with a value well beyond the expected domain of validity. These experiments, coupled with an inversion technique that use the
Eliashberg equations themselves, result in a consistent description of the superconducting state for Pb and other so-called ``strong coupling'' superconductors.

Many systematic deviations from BCS theory have been characterized, for example the gap ratio, $2\Delta_0/(k_BT_c),$\cite{mitrovic84} the specific
heat jump, and many other dimensionless ratios\cite{marsiglio86,marsiglio88a} These have all been reviewed in Ref.~[\onlinecite{carbotte90}], and show
very systematic behaviour as a function of the strong coupling index, $T_c/\omega_{\rm ln}$. On the other hand, when systematics are examined with purely
experimental parameters, the picture is not so clear.\cite{webb15}

\section{Summary and Outlook}

I have provided just a sketch of what we consider the essence of Eliashberg theory --- retardation effects, in the context of a single featureless band. The
generalization of these types of calculations to more complicated scenarios is well documented in a number of places, and have not been reviewed here.
These include order parameter anisotropy, multi-band superconductivity, Berezinskii ``odd-frequency'' pairing, sharply varying electronic density of state,
impurity effects, and so on. These additional complications are increasingly taken into account to understand new classes of compounds that exhibit
superconductivity, such as the hydrides, MgB$_2$, and the pnictides. In some cases, these additional effects have been invoked to explain higher critical
temperatures as well, but for the most part they are motivated by matching theory to experiment.

In its bare form, Fig.~\ref{fig1} presents the possibilities for $T_c$ provided by Eliashberg theory. The conscious decision was made to extend the domain
of coupling strength to unity {\it only} and not beyond, because there are reasons to believe that going beyond this regime is not viable. At the same time,
large values of the characteristic phonon frequency have been used, and this is why the plot extends to beyond $\approx 50$ K for the vertical axis, $T_c$. Are
these values of frequency, together with large values of $\lambda \approx 1$ viable? Probably not, but given these sorts of parameter values,
this is what Eliashberg in its standard form predicts. 

I have also tried to touch on aspects of the theory where more critical scrutiny is possible, by comparing results to those obtained with microscopic models, 
such as the Holstein model. We believe
there are significant difficulties that arise when these comparisons are made. One reaction is to dismiss such comparisons, as the Holstein model
(or the Hubbard model, for that matter) may be regarded as ``toy models,'' possibly fraught with pathologies. However, if the Holstein model is lacking in
some way, it is important to know why, and what other aspect of the electron-phonon interaction (wave-vector dependence?)
is essential to the success of Eliashberg theory. For example, if the super-high-$T_c$ of the hydrides is confirmed to originate in the electron-phonon
interaction, then clearly one or more missing gaps in our understanding of how this happens needs to be filled.

Moreover, as presented here, Eliashberg theory focusses on the superconducting instability, and does not consider other, possibly competing, or potentially
enhancing, instabilities. This possibility has come up as more and more phase diagrams of families of materials exhibit a nearby antiferromagnetic or
charge-density-wave instability, as a function of some tuning parameter (doping, pressure, etc.). It would be desirable to generalize 
Eliashberg theory to be more ``self-regulating,'' and have the theory itself indicate when a competing instability is limiting superconducting $T_c$, for example.

The other aspect that goes hand in hand with the electron-phonon coupling is the direct Coulomb interaction. We cannot claim to understand superconductivity
to the point of having predictive power until we understand the role of Coulomb correlations, and their detrimental (or perhaps favourable?) effects on
pairing. A key advancement has to come in further understanding the role that competitive tendencies or instabilities play in superconductivity. Many of the
modern-day methods (Dynamical Mean Field Theory, for example) seek to address the question of competing interactions. Studies with Quantum Monte Carlo
methods, like the ones mentioned here, will also aid in furthering our understanding of interacting electrons, and similar studies with the now-accessible much
larger lattice sizes would be welcome.

\medskip

\begin{acknowledgments}

This work was supported in part by the Natural Sciences and Engineering
Research Council of Canada (NSERC). I would like to thank the many students and postdoctoral fellows in my own group who have contributed to the work
described here. I also want to thank in particular Jules Carbotte, who first taught me about Eliashberg theory, and Ewald Schachinger, who was instrumental
in teaching me about programming the Eliashberg imaginary axis equations. I also want to thank Sasha Alexandrov, who over the years continued to question the validity of
MIgdal-Eliashberg theory. In the same way, Jorge Hirsch, with whom I have worked a great deal on other matters, has been instrumental in discussions concerning
the validity of the work reviewed here. I also want to thank him for critical comments concerning parts of this review.
Finally, Jules sadly passed away earlier this year (April 5, 2019), and this review is dedicated to his memory. He was a wonderful man,
and I feel extremely fortunate to have first entered the physics world under his guidance.

\end{acknowledgments}
\vskip 0.2in

\noindent{\bf APPENDIX: \ \ Derivation of Eliashberg Theory}
\vskip 0.1in

In this Appendix, we will first outline a derivation of Eliashberg
theory, based on a weak coupling approach. Our primary source for this derivation is Ref.~[\onlinecite{rickayzen65}]. 
Migdal theory of the normal state follows by simply dropping
the anomalous amplitudes in what follows. 

If we know the many-body wave function of system, we can calculate the expectation value for any observable. However,
usually this is something we do not know, and instead we calculate multi-electron Green functions, which are themselves
related to observables. The Green functions are necessarily almost always approximate, and those computed in Eliashberg
theory are no exception. In fact, Eliashberg theory is essentially a mean-field theory, though because of the inherent
frequency dependence in the self energy, it is in many ways a precursor to Dynamical Mean Field Theory.\cite{georges96}

We begin with the definition of the one-electron Green function, defined in momentum space,\cite{mahan00} as a function of imaginary time, $\tau$,
\begin{equation}
G({\bf k},\tau-{\tau^\prime}) \equiv - < T_\tau c_{{\bf k}\sigma}(\tau) c^\dagger_{{\bf k} \sigma}({\tau^\prime}) >,
\label{green_t}
\end{equation}
where ${\bf k}$ is the momentum and $\sigma$ is the spin. The angular
brackets denote a thermodynamic average. We can Fourier-expand this Green function in imaginary frequency:
\begin{eqnarray}
G({\bf k},\tau) = {1\over \beta} \sum_{m=-\infty}^\infty
e^{-i\omega_m \tau} G({\bf k},i\omega_m)
\nonumber \\
G({\bf k},i\omega_m) = \int_0^\beta d\tau G({\bf k},\tau)
e^{i\omega_m \tau}. \label{fourier_green}
\end{eqnarray}
\noindent The frequencies $i\omega_m$ are the Fermion Matsubara
frequencies, given by $i\omega_m = i\pi T(2m-1)$, $m=
0,\pm1,\pm2,...$, where $T$ is the temperature. Because the $c$'s
are Fermion operators, the Matsubara frequencies are {\it odd}
multiples of $i\pi $T. The imaginary time $\tau$ takes on values from 0 to $\beta \equiv {1/(k_B T)}$.

A similar definition holds for the phonon Green function,
\begin{equation}
D({\bf q},\tau-{\tau^\prime}) \equiv -<T_{\tau}A_{\bf
q}(\tau)A_{-{\bf q}}({\tau^\prime})>, \label{phonon_green}
\end{equation}
\noindent where 
\begin{equation}
A_{\bf q}(\tau) \equiv a_{\bf q}(\tau)
\phantom{a} + \phantom{d} a_{ {-}{\bf q}}^\dagger(\tau).
\label{defn_a}
\end{equation}
The Fourier
transform is similar to that given in (\ref{fourier_green}) except
that the Matsubara frequencies are $i\nu_n \equiv i\pi T2n$, $n =
0,\pm1,\pm2,...$ and occur at {\it even} multiples of $i\pi
T$. These are the Boson Matsubara frequencies.

To derive the Eliashberg equations, we follow Ref.~[\onlinecite{rickayzen65}], and use the equation-of-motion method. 
The starting point is the time derivative of Eq. (\ref{green_t}),
\begin{equation}
{\partial \over \partial \tau}G({\bf k},\tau) = -\delta(\tau)
\phantom{a} - \phantom{a} <T_\tau\bigl[H- \mu N,c_{{\bf
k}\sigma}(\tau)\bigr] c^\dagger_{{\bf k}\sigma}(0)>,
\label{deriv1}
\end{equation}
\noindent where we have put ${\tau^\prime} = 0$, without loss of generality.
We use the Hamiltonian (\ref{ham_BKF_mom}), and assume, for the Coulomb
interaction, the simple Hubbard model, $H_{Coul} = U\sum_i
n_{i\uparrow}n_{i\downarrow}$. Including the Coulomb repulsion, the result is repeated here,
\begin{eqnarray}
H & =&  \sum_{{\bf k} \sigma}\epsilon_{\bf k} c^\dagger_{{\bf k}
\sigma} c_{{\bf k} \sigma} \nonumber \\ 
&& + \sum_{\bf q} \hbar \omega_{\bf q} a^\dagger_{\bf q} a_{\bf q} \nonumber \\
&&+ {1 \over \sqrt{N}} \sum_{{\bf k} {\bf k}^\prime \atop \sigma}
g({\bf k},{\bf k}^\prime) \bigl(a_{{\bf k} - {\bf k}^\prime} +
a^\dagger_{-({\bf k} - {\bf k}^\prime)} \bigr) c^\dagger_{{\bf
k}^\prime \sigma} c_{{\bf k} \sigma} \, \, \nonumber \\ && + \, \,
{U \over N} \sum_{\bf k,k^\prime,q} c^\dagger_{{\bf k} \uparrow}
c^\dagger_{{\bf -k + q} \downarrow} c_{{\bf -k^\prime + q}
\downarrow} c_{{\bf k^\prime} \uparrow} , \label{ham_BKF_mom_hubb}
\end{eqnarray}
where the various symbols have already been defined in the text.
We consider only the Green function with $\sigma = \uparrow$;
the commutator in Eq. (\ref{deriv1}) is straightforward and we obtain
\begin{eqnarray}
&&\Biggl({\partial \over \partial \tau} + \epsilon_{{\bf k}} \Biggr)
G_\uparrow ({\bf k},\tau) = \nonumber
\\
&&-\delta(\tau) - {1 \over \sqrt{N}} \sum_{{\bf k}^\prime} g_{{\bf
k} {\bf k}^\prime} <T_\tau A_{{\bf k} - {\bf k}^\prime} (\tau)
c_{{\bf k}^\prime\uparrow}(\tau) c^\dagger_{{\bf k}\uparrow}(0)>
\nonumber \\
&& + {U \over N}\sum_{\bf p p^\prime} <T_\tau c_{{\bf p^\prime - k
+ p} \downarrow}^\dagger (\tau) c_{{\bf p^\prime}\downarrow}(\tau)
c_{{\bf p}\uparrow}(\tau) c^\dagger_{{\bf k}\uparrow}(0)>.
\label{deriv2}
\end{eqnarray}
Various higher order propagators now appear; to
determine them another equation of motion can be written,
which would, in turn, generate even higher order propagators, and this eventually leads to an infinite set of equations with hierarchical
structure. This infinite series is normally truncated at some
point by the process of decoupling, which is simply an
approximation procedure. For example, in (\ref{deriv2}) the
Coulomb term is normally decoupled at this point and becomes
\begin{eqnarray}
\lefteqn{<T_\tau c_{{\bf p^\prime - k + p} \downarrow}^\dagger
(\tau) c_{{\bf p^\prime}\downarrow}(\tau) c_{{\bf
p}\uparrow}(\tau) c^\dagger_{{\bf k}\uparrow}(0)> \rightarrow}
\nonumber \\
&& <T_\tau c_{{\bf p^\prime - k + p} \downarrow}^\dagger (\tau)
c_{{\bf p^\prime}\downarrow}(\tau)> <T_\tau c_{{\bf
p}\uparrow}(\tau) c^\dagger_{{\bf k}\uparrow}(0)>
\rightarrow \nonumber \\
&& -\delta_{{\bf k}{\bf p}}G_\downarrow({\bf p^\prime},0)
G_\uparrow({\bf k},\tau). \label{uterm}
\end{eqnarray}
The case of the electron--phonon term is more difficult; in this case we define a `hybrid' electron/phonon Green function,
\begin{equation}
G_2({\bf k},{\bf k}^\prime,\tau,\tau_1) \equiv <T_\tau A_{{\bf
k}-{\bf k}^\prime}(\tau) c_{{\bf k}^\prime\uparrow}(\tau_1)
c^\dagger_{{\bf k} \uparrow}(0) >, \label{green2}
\end{equation}
\noindent and write out an equation of motion for it.
We simply get
\begin{equation}
{\partial \over \partial \tau}G_2({\bf k},{\bf
k}^\prime,\tau,\tau_1) = -\omega_{{\bf k}-{\bf k}^\prime} <T_\tau
P_{{\bf k}-{\bf k}^\prime}(\tau) c_{{\bf
k}^\prime\uparrow}(\tau_1) c^\dagger_{{\bf k} \uparrow}(0) >,
\label{green2_deriv1}
\end{equation}
\noindent where $P_{\bf q}(\tau) = a_{\bf q}(\tau) - a_{-{\bf
q}}(\tau)$. Taking a second derivative yields
\begin{eqnarray}
&&\lefteqn{\Biggl[ {\partial^2 \over
\partial \tau^2} - \omega_{{\bf k}-{\bf k}^\prime} \Biggr]
G_2({\bf k},{\bf k}^\prime,\tau,\tau_1) =} \nonumber \\
&& \sum_{{\bf k}^{\prime \prime} \sigma}2\omega_{{\bf k}-{\bf
k}^\prime} g_{{\bf k}-{\bf k}^\prime}<T_\tau c^\dagger_{{\bf
k}^{\prime \prime} - {\bf k} + {\bf k}^\prime \sigma}(\tau)
c_{{\bf k}^{\prime \prime}\sigma}(\tau) c_{{\bf k}^\prime
\uparrow} (\tau_1)c^\dagger_{{\bf k} \uparrow}(0)>.\nonumber \\
\label{green2_deriv2}
\end{eqnarray}
\noindent At this point we {\it do not simply} decouple the last line of Eq.~(\ref{green2_deriv2}). We first need to
take the phonon propagator into account, and the standard procedure is to use the ``non-interacting'' phonon
propagator. The adjective ``non-interacting'' is in quotes because part of the philosophy of proceeding in this way was a desire
to {\it not} compute corrections to the phonon propagator, because the information going into this part of the calculations (e.g. the phonon spectral function)
was going to come from experiment. Coming from experiment means that nature ``had already done the calculation,'' and we did not want to double count.
Clearly, if the purpose of this calculation is to compare to Quantum Monte Carlo calculations where this is {\it not} the case, then something different should be
done, and this is what motivated the {\it renormalized} Migdal-Eliashberg calculations of Refs.~[\onlinecite{marsiglio90,marsiglio91}]. 

For now, we proceed with the standard Eliashberg calculations. The equation of motion for the non-interacting phonon propagator is
\begin{equation}
\Biggl( {\partial^2 \over \partial \tau^2} - \omega_{\bf q}^2
\Biggr) D({\bf q},\tau - {\tau^\prime}) = 2\omega_{\bf q}
\delta(\tau - {\tau^\prime}). \label{phonon_prop}
\end{equation}
\noindent Inserting this expression into Eq. (\ref{green2_deriv2}) then yields
\begin{eqnarray}
&&G_2({\bf k},{\bf k}^\prime,\tau,\tau)  =  {1 \over N} \sum_{{\bf
k}^{\prime \prime}\sigma}\int_0^\beta d{\tau^\prime} g_{\bf k^{\prime \prime},{\bf k^{\prime \prime}+k-k^\prime}} %g_{{\bf k} {\bf k}^\prime}
D({\bf k} - {\bf k}^\prime, \tau - {\tau^\prime})
\nonumber
\\
&& \times <T_\tau c^\dagger_{{\bf k}^{\prime \prime} \sigma}({\tau^\prime}) c_{{\bf k^{\prime \prime} + k - k^\prime} \sigma}({\tau^\prime}) 
% c^\dagger_{{\bf k}^{\prime \prime} - {\bf k} + {\bf k}^\prime \sigma}({\tau^\prime}) c_{{\bf k}^{\prime \prime}\sigma}({\tau^\prime}) 
c_{{\bf k}^\prime \uparrow}(\tau)c^\dagger_{{\bf k} \uparrow}(0)>, \label{green_deriv3}
\end{eqnarray}
\noindent where now $\tau_1$ has been set equal to $\tau$ as is
required in (\ref{deriv2}). It is important that this be done only {\it after} applying Eq.~(\ref{phonon_prop}). 
The result can be substituted into Eq.~(\ref{deriv2}), and then Fourier transformed
(from imaginary time to imaginary frequency). Before doing this however, we recall that Gor'kov\cite{gorkov58} realized the important role of the
so-called Gor'kov anomalous amplitude, in the Wick
decomposition \cite{mahan00} of the various two--particle Green
functions encountered above. We therefore have to account for these in addition to the pairing of fermion operators used in Eq.~(\ref{uterm}).

The anomalous amplitudes are defined to be
\begin{equation}
F({\bf k},\tau) \equiv -<T_\tau c_{{\bf k} \uparrow}(\tau)c_{-{\bf
k} \downarrow}(0)> \label{f_amplitude}
\end{equation}
\noindent and
\begin{equation}
\bar{F}({\bf k},\tau) \equiv -<T_\tau c^\dagger_{-{\bf k}
\downarrow}(\tau) c^\dagger_{{\bf k} \uparrow}(0)>.
\label{fbar_amplitude}
\end{equation}
Now we need to repeat the same steps as above with $F$ and $\bar{F}$
as we did with $G$. Skipping the intermediate steps, the result is an equation analogous to Eq.~(\ref{deriv2})
\begin{eqnarray}
&&\Biggl({\partial \over \partial \tau} - \epsilon_{{\bf k}} \Biggr)
\bar{F} ({\bf k},\tau) = \nonumber
\\
&& - {1 \over \sqrt{N}} \sum_{{\bf k}^\prime} g_{{\bf
-k^\prime, -k}} <T_\tau A_{{\bf k} - {\bf k}^\prime} (\tau)
c^\dagger_{{\bf -k}^\prime\downarrow}(\tau) c^\dagger_{{\bf k}\uparrow}(0)>
\nonumber \\
&& + {U \over N}\sum_{\bf k^{\prime \prime},q} 
%<T_\tau c_{{\bf p^\prime - k+ p} \downarrow}^\dagger (\tau) c_{{\bf p^\prime}\downarrow}(\tau) c_{{\bf p}\uparrow}(\tau) c^\dagger_{{\bf k}\uparrow}(0)>,
<T_\tau c^\dagger_{{\bf k^{\prime \prime}} \uparrow} (\tau) c^\dagger_{{\bf -k^{\prime \prime} + q}\downarrow}(\tau) c_{{\bf k+q}\uparrow}(\tau) c^\dagger_{{\bf k}\uparrow}(0)>, \nonumber \\
\label{derivf2}
\end{eqnarray}
and similarly for the function $F$. This leads to the need for another `hybrid' electron/phonon anomalous Green function,
\begin{equation}
\bar{F}_2({\bf k},{\bf k}^\prime,\tau,\tau_1) \equiv <T_\tau A_{{\bf
k}-{\bf k}^\prime}(\tau) c^\dagger_{{\bf -k}^\prime \downarrow}(\tau_1)
c^\dagger_{{\bf k} \uparrow}(0) >, \label{greenf2}
\end{equation}
and, following the same process as for the regular Green function, we find
\begin{eqnarray}
&&\bar{F}_2({\bf k},{\bf k}^\prime,\tau,\tau)  =  {1 \over N} \sum_{{\bf
k}^{\prime \prime}\sigma}\int_0^\beta d{\tau^\prime} g_{\bf k^{\prime \prime},{\bf k^{\prime \prime}+k-k^\prime}} %g_{{\bf k} {\bf k}^\prime}
D({\bf k} - {\bf k}^\prime, \tau - {\tau^\prime})
\nonumber
\\
&& \times <T_\tau c^\dagger_{{\bf k}^{\prime \prime} \sigma}({\tau^\prime}) c_{{\bf k^{\prime \prime} + k - k^\prime} \sigma}({\tau^\prime}) 
% c^\dagger_{{\bf k}^{\prime \prime} - {\bf k} + {\bf k}^\prime \sigma}({\tau^\prime}) c_{{\bf k}^{\prime \prime}\sigma}({\tau^\prime}) 
c^\dagger_{{\bf -k}^\prime \downarrow}(\tau)c^\dagger_{{\bf k} \uparrow}(0)>, \label{greenf_deriv3}
\end{eqnarray}
where again $\tau_1$ has been set equal to $\tau$ {\it after} applying Eq.~(\ref{phonon_prop}). 

The Fourier definitions of the anomalous Green function are the same as Eq.~(\ref{fourier_green}):
\begin{eqnarray}
\bar{F}({\bf k},\tau) = {1\over \beta} \sum_{m=-\infty}^\infty
e^{-i\omega_m \tau} \bar{F}({\bf k},i\omega_m)
\nonumber \\
\bar{F}({\bf k},i\omega_m) = \int_0^\beta d\tau \bar{F}({\bf k},\tau)
e^{i\omega_m \tau},
\label{fourier_greenf}
\end{eqnarray}
and similarly for $F$.
In frequency space one finds that two self energies naturally arise,
\bwt

\begin{equation}
\Sigma({\bf k},i\omega_m) = -{1 \over N \beta} \sum_{{\bf k^\prime},m^\prime} g_{\bf kk^\prime}g_{\bf k^\prime k}
D({\bf k - k^\prime},i\omega_m - i\omega_{m^\prime}) G({\bf k^\prime},i\omega_{m^\prime}),
\label{sigma}
\end{equation}

\begin{equation}
\phi({\bf k},i\omega_m) = -{1 \over N \beta} \sum_{{\bf k^\prime},m^\prime} g_{\bf k^\prime k}g_{\bf -k^\prime -k}
D({\bf k - k^\prime},i\omega_m - i\omega_{m^\prime}) F({\bf k^\prime},i\omega_{m^\prime}).
\label{sigma}
\end{equation}
%\ewt
One can show that $g_{\bf k^\prime k} = g^\ast_{\bf k k^\prime}$; normally one expects a similar relation with negative wave vectors, and we assume it in what follows.
These equations are then written self-consistently and lead to Eqs.~(\ref{ga1}-\ref{ga4}) once Eq.~(\ref{even_odd}) is used.
\ewt

%\appendix
%
%\section{probably not needed}


\begin{thebibliography}{99}
\bibitem{remark1} We could have elected to keep the area, $\int_0^\infty d\nu \alpha^2F(\nu)$ constant as well, for example.
\bibitem{ewald} There are undoubtedly references that describe this procedure in more detail, but I learned it primarily from a mentor at the time,
Ewald Schachinger, who was a frequent visitor at McMaster University during the course of my PhD.
\end{thebibliography}
\end{document}